\documentclass[12pt, draftclsnofoot, onecolumn]{IEEEtran}
\ifCLASSINFOpdf
\else
\fi
\usepackage{amsthm,amsmath,amssymb}
\usepackage{mathrsfs}
\usepackage{bm}
\usepackage{graphicx}
\usepackage{stfloats}
\usepackage{cite}
\usepackage{multirow}
\usepackage{rotating}
\usepackage{color}
\usepackage{makecell}

\begin{document}
%
\title{Changeable Rate and Novel Quantization for CSI Feedback Based on Deep Learning}
%
%
%

\author{Xin~Liang,
        Haoran~Chang,
        Haozhen~Li,
        Xinyu~Gu,~\IEEEmembership{Member,~IEEE,}
        and~Lin~Zhang,~\IEEEmembership{Member,~IEEE}
\thanks{X. Liang, H. Chang, H. Li and X. Gu are with the School of Artificial Intelligence,
Beijing University of Posts and Telecommunications, Beijing 100876, China 
(e-mail: \{liangxin, changhaoran, lihaozhen, guxinyu\}@bupt.edu.cn)}
\thanks{L. Zhang is with the School of Information and Communication Engineering, 
Beijing Information Science and Technology University, Beijing 100085, China and 
also with the School of Artificial Intelligence, Beijing University of Posts and Telecommunications,
Beijing 100876, China (e-mail: zhanglin@bupt.edu.cn)}
}

%
%

\markboth{}
{Shell \MakeLowercase{\textit{et al.}}: Bare Demo of IEEEtran.cls for IEEE Journals}
%



\maketitle

\begin{abstract}
  Deep learning (DL)-based channel state information (CSI) feedback improves the capacity and energy efficiency of massive multiple-input multiple-output (MIMO) systems in frequency division duplexing mode. However, multiple neural networks with different lengths of feedback overhead are required by time-varying bandwidth resources. The storage space required at the user equipment (UE) and the base station (BS) for these models increases linearly with the number of models. In this paper, we propose a DL-based changeable-rate framework with novel quantization scheme to improve the efficiency and feasibility of CSI feedback systems. This framework can reutilize all the network layers to achieve overhead-changeable CSI feedback to optimize the storage efficiency at the UE and the BS sides. Designed quantizer in this framework can avoid the normalization and gradient problems faced by traditional quantization schemes. Specifically, we propose two DL-based changeable-rate CSI feedback networks CH-CsiNetPro and CH-DualNetSph by introducing a feedback overhead control unit. Then, a pluggable quantization block (PQB) is developed to further improve the encoding efficiency of CSI feedback in an end-to-end way. Compared with existing CSI feedback methods, the proposed framework saves the storage space by about 50\% with changeable-rate scheme and improves the encoding efficiency with the quantization module.
\end{abstract} 


%
\IEEEpeerreviewmaketitle

\section{Introduction}
%
%
%
%
\IEEEPARstart{M}{assive} multiple-input multiple-output (MIMO) has been proved to be a promising technology for beyond 5G and next wireless communication systems \cite{ref1,ref2,ref3,ref4}. By deploying large-scale antenna arrays, the base station (BS) can achieve high downlink throughput and low interference \cite{ref5,ref6}. The above benefit requires instantaneous and accurate downlink channel state information (CSI) at the BS side \cite{ref7}. In time division duplexing (TDD) mode, uplink and downlink work in the same frequency band but different time slots. Thus, BS can estimate downlink CSI utilizing reciprocity from uplink CSI. However, in widely used frequency division duplexing (FDD) systems, downlink CSI is hard to infer because of the obscure reciprocity between the uplink and downlink frequency bands.

Existing FDD MIMO systems often use direct quantization approach for downlink CSI feedback. Specifically, the user equipment (UE) estimates the current time downlink CSI with the pilot sent from the BS firstly. Then, the UE quantifies CSI in the form of precoding matrix index (PMI) and reports PMI to the BS using feedback link \cite{ref8}. However, with the increasing number of antennas in massive MIMO communication system, linearly growing feedback overhead occupies excessive spectrum which is not acceptable. To tackle this problem, it is necessary to find a more efficiency method to compress and sense CSI. Compressive sensing (CS)-based algorithms compress and reconstruct CSI \cite{ref9, ref10} under the assumption of the sparsity of channel in a certain domain. However, the sparsity assumption of channel leads to limited performance of CS-based approaches. Moreover, because of the time-varying nature of the channel, CSI is time sensitive. The iterative algorithms based on CS are time-consuming which reduce the performance gain provided by downlink CSI.

Deep learning (DL) has achieved a great success in many fields \cite{ref11,ref12, ref13, ref14, ref15}. Driven by large datasets, the algorithms based on DL have been proved to be able to provide a reliable solution to the problems that are difficult to model. Through forward and back-propagation, neural networks also have a low time delay.

In the field of wireless communications, the CSI matrix can be regarded as an image because of the correlation of adjacent elements. Thus, it is possible to process CSI matrix using approaches based on DL, e.g., for channel estimation \cite{ref16,ref17}, feedback \cite{ref18,ref19,ref20,ref21,ref22,ref23}, signal detection \cite{ref24}, channel modeling \cite{ref25} and scenario identification \cite{ref26}.

The authors of \cite{ref18}, for the first time, introduce a DL-based framework for CSI feedback called CsiNet. CsiNet employs the autoencoder structure consisting of encoder and decoder. Encoder is deployed at the UE side to sense and compress CSI into a low rank codeword vector. Then, this codeword is reported to the BS using feedback link. Finally, decoder deployed at the BS side reconstructs codeword vector to obtain the original CSI. CsiNet has been demonstrated to have better reconstruction accuracy and lower time consumption compared with CS-based schemes. Next, researchers develop a series of architectures to exploit the correlations of wireless channel to improve CSI feedback and reconstruction accuracy. DualNet is proposed in \cite{ref19}, which exploits bi-directional channel implicit reciprocity in DL to improve downlink CSI reconstruction accuracy with the help of uplink CSI. Using LSTM architecture \cite{ref20}, time correlation is utilized to improve the accuracy of CSI feedback. In \cite{ref21}, the spatial correlation of multiple users is considered to reduce CSI feedback overhead. Some works focus on excavating the abilities of encoder and decoder to achieve high performance. After investigating the characteristics of CSI, the authors of \cite{ref22} establish a guideline for CSI feedback network designing and proposed an advanced architecture based on autoencoder, named CsiNet+. CsiNet+ extracts deeper channel features and shows competitive performance. The architectures of CsiNet and DualNet are optimized in \cite{ref23}, and their advanced counterparts named CsiNetPro and DualNetSph are proposed, to produce more efficient CSI codewords and achieve more accurate CSI reconstruction accuracy.

Among the aforesaid DL-based CSI feedback schemes, most of them are designed through autoencoder framework to compress CSI. To make neural network run properly, the compressed CSI, i.e., the codeword, is required to have a fixed length. However, CSI feedback overhead is subject to change according to the available bandwidth resource and the reconstruction accuracy requirement of CSI. 3GPP TS 38.214 \cite{ref8} stipulates that, according to the bandwidth resource and the settings of feedback system, PMI can be fed back with a variable length. The rest of PMI information which exceeds the limitation of feedback overhead will be discarded directly. If a part of the codeword is truncated simply in the DL-based CSI feedback approach, autoencoder will work abnormally. he authors of \cite{ref22} propose two architectures called SM-CsiNet+ and PM-CsiNet+, which support a 4-option selectable CSI feedback overhead by training multiple decoders at the BS side. Similar to \cite{ref22}, a serial compression scheme SALDR is developed in \cite{SALDR} to support at most 4-option feedback overhead in DL-based CSI feedback networks. However, such few options still cannot meet the actual application requirements. Moreover, the number of decoders increases with the number of options linearly, which occupies precious storage resources. Thus, designing a simple and feasible framework supporting fine-grained changeable CSI feedback overhead is urgently needed.

Meanwhile, existing works focus on improving the CSI reconstruction accuracy of neural networks, but most of them ignore the impact of quantization operations. The quantization of codewords can improve the encoding efficiency but introduces quantization noise. Therefore, finding an efficient quantizer with minimized quantization noise is important to improve the CSI feedback system performance. Due to the fact that quantization operation is not differentiable, quantization cannot be directly conducted in the back-propagation process of the neural network training. An offset module is developed in \cite{ref22} to reduce the quantization distortion and a dedicated training strategy is established to avoid the gradient problem. The authors of \cite{ref27} design an end-to-end CSI feedback framework with quantization operations where the quantized gradient is forced to the constant one. However, the above approximate quantizers do not fully consider the behavior of the quantizers in DL-based tasks and operate in fragile ways that affect the convergence of the neural networks. To get the optimal fitting solution, the quantization network should be globally optimized and a proper back-propagation gradient is required.

To deal with the above storage and encoding efficiency challenges, we propose a DL-based changeable-rate CSI feedback scheme, which improves the storage efficiency and reduces the quantization noise. Specifically, we first propose two changeable-rate CSI feedback networks for variable feedback overhead to save the storage space of the UE and BS. Then, we develop a novel quantizer to minimize the noise introduced by quantization operation. Finally, we analyze the performance of the proposed networks and discuss the mechanisms of changeable-rate CSI feedback and the proposed quantization scheme.

The main contributions of this work are summarized as follows:

\begin{itemize}
  \item To improve the efficiency and feasibility of CSI feedback systems, we propose a universal DL-based changeable-rate CSI feedback framework with a novel quantizer. This framework reutilizes all the neural network layers to extract the general features under different feedback overheads to achieve variable length coding of CSI, which can reduce the storage overhead of models at both the UE side and the BS side. The quantizer can improve the encoding efficiency and minimize the impact of quantization noise.
  \item Two changeable-rate CSI feedback networks, CH-CsiNetPro and CH-DualNetSph, are proposed by introducing a feedback overhead control unit. With the proposed training strategy, CH-CsiNetPro and CH-DualNetSph can reconstruct CSI from the length-changeable feedback codewords with only one transmitter (encoder) at the UE and one receiver (decoder) at the BS. Compared with positioning multiple length-fixed CSI feedback networks, CH-CsiNetPro and CH-DualNetSph can save the storage space and keep robust CSI recovery accuracy. For a typical setting of length-fixed CSI feedback scheme, it is possible to reduce the storage space by about 50\% while not increasing the amount of floating-point operations (FLOPs) needed at both the UE and the BS sides.
  \item We utilize a bounded mapping function and design an approximate gradient for the proposed quantizer named pluggable quantization block (PQB). PQB can avoid the normalization and gradient problems faced by existing quantization schemes and optimize CSI feedback networks in an end-to-end way. Experiment results show that PQB achieves better CSI reconstruction accuracy compared with existing quantization schemes. Combined with the introduced changeable-rate CSI feedback networks, we further propose two joint optimization networks named CH-CsiNetPro-PQB and CH-DualNetSph-PQB to improve the storage and encoding efficiency of CSI feedback system at the same time.
  \item We finally analyze the mechanism of changeable-rate CSI feedback and evaluates the proposed quantization frameworks using information entropy, which provides a guideline for future researches on DL-based variable length coding and end-to-end quantization for CSI. The investigations of the efficiency and reconstruction accuracy improvement are discussed.
\end{itemize}

The rest of this paper is organized as follows. In Section \uppercase\expandafter{\romannumeral2}, we formulate the system model. Section \uppercase\expandafter{\romannumeral3} introduces a changeable-rate CSI feedback scheme based on DL. To further show the efficiency of the introduced scheme, we propose two changeable-rate feedback networks, CH-CsiNetPro and CH-DualNetSph. Section \uppercase\expandafter{\romannumeral4} introduces a novel quantization module to end-to-end optimize the encoding efficiency of CSI feedback architectures. Finally, we design the experiment for the proposed networks, provide numerical performance results and discuss the efficiency and performance improvement of the designed changeable-rate scheme and quantization module for DL-based CSI feedback networks in Section \uppercase\expandafter{\romannumeral5}. Section \uppercase\expandafter{\romannumeral6} concludes the paper.

\section{System Model}

In this section, we introduce the wireless communication system of massive MIMO orthogonal frequency division multiplexing (OFDM). Then, we discuss CSI compression, quantization, feedback and reconstruction mechanism based on DL.

\subsection{Massive MIMO OFDM System}
Consider a single-cell massive MIMO OFDM system, $N_t\gg1$ transmitting antennas are deployed at the BS side and a single receiving antenna is positioned at the UE side. There are $N_s$ subcarriers adopted in this system. Thus, the downlink received signal at the $i$-th subcarrier is described as
\begin{equation}
  y_d^i={\tilde{\mathbf{h}}{_{d}^{i}}}^H\mathbf{v}_t^ix_d^i+n_d^i, \label{eq_1} 
\end{equation}

\noindent where ${\tilde{\mathbf{h}}}_d^i\in\mathbb{C}^{N_t\times1}$ denotes the downlink channel frequency response vector of the $i$-th subcarrier, $\mathbf{v}_t^i\in\mathbb{C}^{N_t\times1}$ represents the transmitted precoding vector, $x_d^i\in\mathbb{C}$ is the downlink sent symbol and $n_d^i\in\mathbb{C}$ denotes the additive noise. $\left(\cdot\right)^H$ denotes conjugate transpose. The BS can calculate transmitted precoding vector $\mathbf{v}_t^i$ once the downlink CSI vector ${\tilde{\mathbf{h}}}_d^i$ has been obtained. The uplink received signal at the $i$-th subcarrier is
\begin{equation}
  y_u^i={\mathbf{v}_r^i}^H{\tilde{\mathbf{h}}}_u^ix_u^i+{\mathbf{v}_r^i}^Hn_u^i, \label{eq_2} 
\end{equation}

\noindent where $\mathbf{v}_r^i\in\mathbb{C}^{N_t\times1}$ denotes the receiving beamformer of the $i$-th subcarrier, ${\tilde{\mathbf{h}}}_u^i\in\mathbb{C}^{N_t\times1}$ denotes the uplink channel vector, $x_u^i\in\mathbb{C}$ is the uplink sent symbol and $n_u^i\in\mathbb{C}$ is the additive noise. The downlink channel vectors of $N_s$ subcarriers stacked in the frequency domain is ${\tilde{\mathbf{H}}}_d=[{\tilde{\mathbf{h}}}_d^1,{\tilde{\mathbf{h}}}_d^2,\ldots,{\tilde{\mathbf{h}}}_d^{N_s}]^H\in\mathbb{C}^{N_s\times N_t}$. Similarly, the uplink CSI matrix in the spatial-frequency is denoted as ${\tilde{\mathbf{H}}}_u=[{\tilde{\mathbf{h}}}_u^1,{\tilde{\mathbf{h}}}_u^2,\ldots,{\tilde{\mathbf{h}}}_u^{N_s}]^H\in\mathbb{C}^{N_s\times N_t}$.

In FDD mode, the UE needs to feed downlink CSI matrix back to the BS to calculate transmitted precoding vector. However, the total number of real-valued feedback parameters is $2N_sN_t$. Undue feedback payload makes band resource exhausted. To exploit the sparsity of CSI matrix in the angular-delay domain, $\tilde{\mathbf{H}}$ is transformed from the spatial-frequency domain to the angular-delay domain using 2-D discrete Fourier transform (DFT), i.e.,
\begin{equation}
  \mathbf{H}=\mathbf{F}_\text{d}\tilde{\mathbf{H}}\mathbf{F}_\text{a}^H, \label{eq_3} 
\end{equation}

\begin{figure*}[htbp]
  \centerline{\includegraphics[scale=0.55]{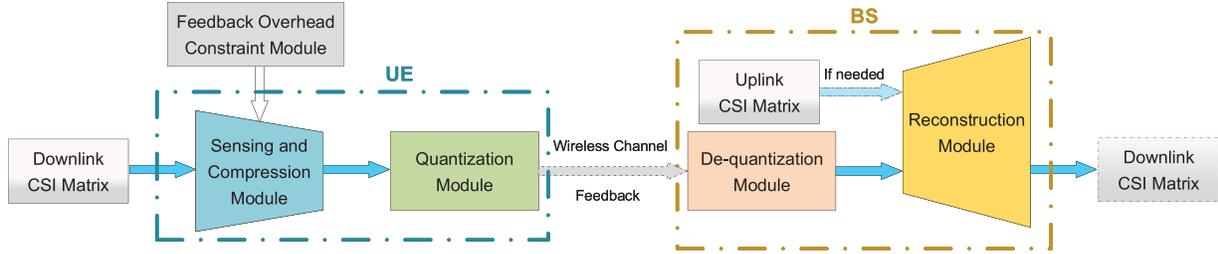}}
  \caption{The proposed DL-based changeable-rate CSI feedback framework with quantization operation.}
\label{fig1} 
\end{figure*}

\noindent where $\mathbf{F}_\text{d}$ and $\mathbf{F}_\text{a}$ are $N_s\times N_s$ and $N_t\times N_t$ DFT matrix, respectively. After 2-D DFT, the elements of $\mathbf{H}$ are near zero except the first ${\tilde{N}}_s$ rows in the delay domain. Thus, only the first ${\tilde{N}}_s$ rows of $\mathbf{H}$ are retained, and the truncated CSI matrix has most information of $\mathbf{H}$. Let $\mathbf{H}_d$ and $\mathbf{H}_u$ be the truncated CSI matrix of downlink and uplink in the angular-delay domain, respectively. The total number of real-valued feedback parameters of $\mathbf{H}_d$ decreases to $2{\tilde{N}}_sN_t$, which is still a huge overhead for massive MIMO system. Although the elements of $\mathbf{H}_d$ retain most information of original CSI matrix, $\mathbf{H}_d$ is still sparse and has local relevance. Therefore, it is possible to further reduce the feedback overhead by compression using neural networks.


\subsection{DL-based CSI Feedback}
An autoencoder architecture is applied to compress, report and reconstruct CSI matrix. Different from existing works, a scenario with changeable feedback overhead is taken into consideration. As shown in Fig. \ref{fig1}, we propose a DL-based changeable-rate CSI feedback framework with quantization operation. There are sensing, compression and quantization modules at the UE side, de-quantization and reconstruction modules at the BS side. The UE compresses downlink CSI matrix into a codeword. Because the limitation of time-varying bandwidth resource  places constraints on the feedback overhead, the length of codeword vector is subject to change. Then, the codeword is sent to quantization module and fed back to the BS as a bit stream. De-quantization module recovers the codeword from the received bit stream. After that, the BS reconstructs CSI matrix from codeword with the help of uplink CSI matrix (if needed).

The codeword sensed and compressed from $\mathbf{H}_d$ using the encoder $f_\text{en}\left(\cdot\right)$ which is denoted as
\begin{equation}
  \mathbf{s}_n=f_\text{en}\left(\mathbf{H}_d,n\right), \label{eq_4} 
\end{equation}
where $n$ is the length of codeword under the feedback overhead constraint. The quantization and de-quantization of codeword $\mathbf{s}_n$ which is denoted as
\begin{equation}
  {\hat{\mathbf{s}}}_n=f_\text{de-quan}\left(f_\text{quan}\left(\mathbf{s}_n\right)\right), \label{eq_5} 
\end{equation}
where $f_\text{quan}\left(\cdot\right)$ and $f_\text{de-quan}\left(\cdot\right)$ are the quantization and de-quantization function, respectively.

The process of recovering CSI matrix from ${\hat{\mathbf{s}}}_n$ using the decoder $f_\text{de}\left(\cdot\right)$ is denoted as the follows
\begin{equation}
  {\hat{\mathbf{H}}}_d=f_\text{de}\left({\hat{\mathbf{s}}}_n\right). \label{eq_6} 
\end{equation}

\begin{figure}[htbp]
  \centerline{\includegraphics[scale=0.74]{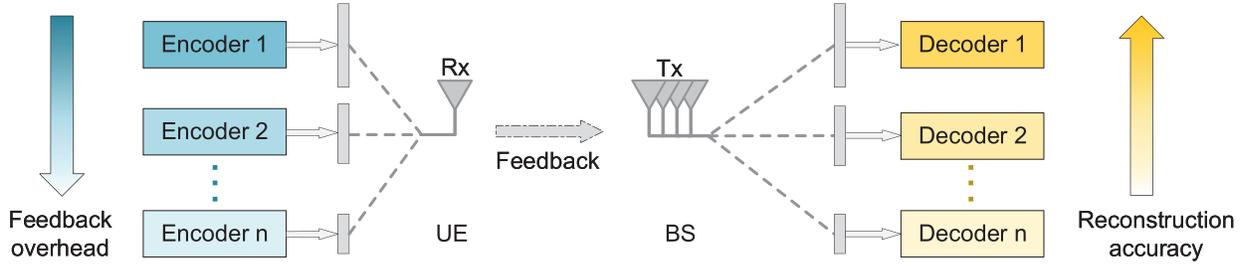}}
  \caption{Illustration of multiple neural networks working under different feedback overhead. Due to the bandwidth resource constraint, the UE needs to select an appropriate encoder and the corresponding decoder to compress CSI to a certain length. There is a trade-off between feedback overhead and reconstruction accuracy.}
  \label{fig2} 
\end{figure}

If uplink CSI matrix $\mathbf{H}_u$ is used as an auxiliary, equation \eqref{eq_6} becomes
\begin{equation}
  {\hat{\mathbf{H}}}_d=f_\text{de}\left({\hat{\mathbf{s}}}_n,\mathbf{H}_u\right). \label{eq_7} 
\end{equation}

The mean-squared error (MSE) is utilized as the objective function for the training of neural network, which is formulated as minimizing
\begin{equation}
  \text{MSE}=\frac{1}{N}\sum_{k=1}^N\Vert{\mathbf{H}_d^k-\hat{\mathbf{H}}_d^k}\Vert_2^2, \label{eq_8} 
\end{equation}
where $N$ is the total number of training samples of the neural network and $k$ is the index of the sample.

\section{DL-based Changeable-Rate CSI Feedback}

Existing DL-based CSI feedback frameworks, e.g. CsiNetPro and DualNetSph\cite{ref23}, have shown to have a great performance on CSI sensing, compressing and recovering. However, once the neural network is trained, the input data size and the output data size are fixed for each layer. Therefore, the length of feedback codeword is immutable. As shown in Fig. \ref{fig2}, multiple neural networks trained with different feedback overhead options need to be stored at the BS and the UE to deal with the constraint of variable bandwidth resource and feedback accuracy. Redundant structures occupy excessive storage space and make the system inefficient.

In this section, CsiNetPro and DualNetSph are taken as examples to show the number of parameters of neural networks with different feedback codeword overhead. Then, a changeable-rate CSI feedback scheme is proposed to improve the efficiency of the system.

\subsection{Architectures of DL-based CSI feedback}
As described in\cite{ref23}, CsiNetPro senses and compresses the complex CSI matrix in Cartesian coordinate system. At the UE side, four convolutional layers utilize $7\times7$ kernels to generate 16, 8, 4 and 2 feature maps, which extract the features of CSI matrix. Then, after flattened into a vector, an $M$-element fully connected (FC) layer is used to reduce dimension and generate the codeword $\mathbf{s}_M$. The decoder, at the BS side, firstly use an FC layer to restore the original dimension of CSI matrix, and the output matrix is sent to a series of convolutional layers of $7\times7$ kernel with 16, 8, 4 and 2 feature maps, sequentially, to reconstruct the real and imaginary parts of CSI matrix. Different from CsiNetPro, DualNetSph exploits the magnitude reciprocity of bidirectional channel to improve the reconstruction accuracy of CSI in polar coordinate system. The compression and feedback process can be separated into two parts. The magnitude of CSI matrix is fed into four convolutional layers of $7\times7$ kernel to generate 16, 8, 4 and 1 feature maps, sequentially, at the encoder. Then, the feature map is reshaped into a vector and compressed by an FC layer with $M$ elements. At the BS side, the decoder uses an FC layer to improve the dimension of the codeword before compression. Then, uplink CSI magnitude is leveraged as an auxiliary input to help reconstruct downlink CSI. The combination of codeword and uplink CSI magnitude is sent to a series of convolutional layers of $7\times7$ kernel to generate 16, 8, 4 and 1 feature maps to recovered the original CSI magnitude matrix. To improve the efficiency of CSI feedback, DualNetSph adopts a magnitude dependent phase quantization (MDPQ) approach to quantize the phase of CSI matrix. MDPQ uses finer quantization steps to quantize the phase with larger magnitude and vice versa.

To evaluate the complexity of neural networks, we follow the widely-used settings in the DL-based CSI feedback systems \cite{ref18,ref19,ref20,ref21,ref22,ref23} and set the number of transmitting antennas $N_t=32$. The number of subcarriers is set to $N_s=1024$, and the first ${\tilde{N}}_s=32$ rows are retained in the truncated CSI matrix in the angular-delay domain. Therefore, the size of complex CSI matrix is $N_t\times{\tilde{N}}_s$, i.e., $32\times32$. The lengths of the vector before sent to the FC layers for dimension reduction are $2048$ and $1024$ for CsiNetPro and DualNetSph, respectively.

\begin{table}[]
  \centering
  \caption{The Numbers of Trainable Parameters of CsiNetPro and DualNetSph with Different Feedback Codeword Lengths $M$}
\label{table1}
  \setlength{\tabcolsep}{1.15mm}{
    \begin{tabular}{c|c|c|c|c|c|c}
      \hline
                                  & $M$                                                    & 32      & 64      & 128     & 256       & 512       \\ \cline{2-7} 
      \multirow{3}{*}{CsiNetPro}  & \begin{tabular}[c]{@{}c@{}}Encoder\\ (at the UE)\end{tabular} & 75,654  & 141,222 & 272,358 & 534,630   & 1,059,174 \\ \cline{2-7} 
                                  & \begin{tabular}[c]{@{}c@{}}Decoder\\ (at the BS)\end{tabular} & 77,606  & 143,142 & 274,214 & 536,358   & 1,060,646 \\ \cline{2-7} 
                                  & Total                                                         & 153,260 & 284,364 & 546,572 & 1,070,988 & 2,119,820 \\ \hline
                                  & $M$                                                    & 16      & 32      & 64      & 128       & 256       \\ \cline{2-7} 
      \multirow{3}{*}{DualNetSph} & \begin{tabular}[c]{@{}c@{}}Encoder\\ (at the UE)\end{tabular} & 25,505  & 41,905  & 74,705  & 140,305   & 271,505   \\ \cline{2-7} 
                                  & \begin{tabular}[c]{@{}c@{}}Decoder\\ (at the BS)\end{tabular} & 27,233  & 43,617  & 76,385  & 141,921   & 272,993   \\ \cline{2-7} 
                                  & Total                                                         & 52,738  & 85,522  & 151,090 & 282,226   & 544,498   \\ \hline
      \end{tabular}
  }
  \end{table}

\subsection{Complexity Analysis}

Neural networks contain a huge number of trainable parameters to fit the dataset. For the aforesaid DL-based CSI feedback architectures, there are trainable parameters in the convolutional layer, FC layer and batch normalization layer. The number of trainable parameters of batch normalization layer in our scheme is a fixed value $64$. The trainable parameters calculation formulas of the convolutional layer and FC layer are defined as follows:
\begin{equation}
  \begin{split}
  P_C&=\left(C_{in}\times K^2+1\right)C_{out},\\
  P_F&=F_{out}(F_{in}+1),
  \end{split}
  \label{eq_9} 
\end{equation}

\noindent where $P_C$ and $P_F$ denote the numbers of parameters of the convolutional layer and FC layer, respectively. $C_{in}$ and $C_{out}$ are the numbers of input and output convolutional feature maps. $K$ is the size of convolutional kernel. $F_{in}$ and $F_{out}$ denote the numbers of input and output elements of FC layer, respectively. 


Table \ref{table1} respectively lists the numbers of trainable parameters $P_B+P_C+P_F$ for CsiNetPro and DualNetSph with different feedback codeword length $M$. The number of trainable parameters increases with the increasing of feedback overhead $M$. Training multiple CSI feedback networks with different feedback codeword lengths will consume huge storage space at both the UE and the BS side, which is infeasible. A low-complexity and feasible CSI feedback framework supporting length-changeable feedback codeword is needed to improve the storage efficiency for massive MIMO system.

\begin{figure*}[]
  \centerline{\includegraphics[scale=0.74]{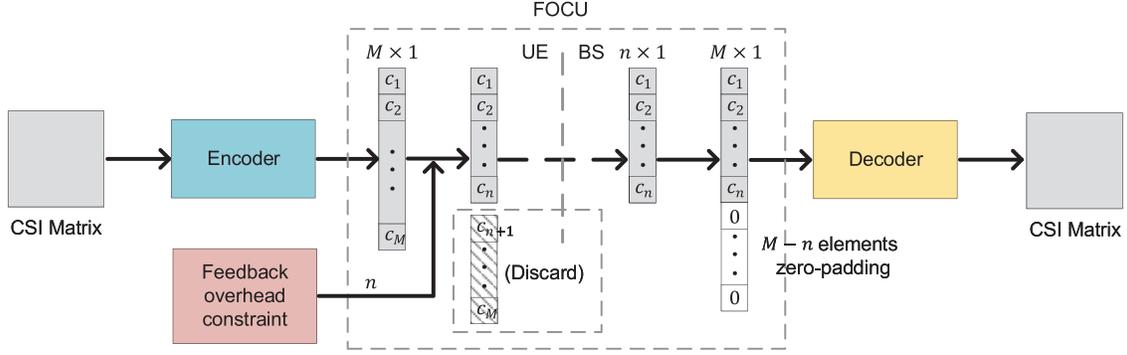}}
  \caption{The illustration of the proposed feedback overhead control unit (FOCU). FOCU firstly discards the last $M-n$ elements of the codeword generated by encoder under the feedback overhead constraint. Then, FOCU zero-pads $M-n$ zeros to the truncated codeword so that the training of the neural network is feasible.}
  \label{fig4} 
\end{figure*}

\subsection{Feedback Overhead Control Unit}
Due to the attribute of the FC layer, its operation is length-fixed. Therefore, the degree of dimension reduction of the CSI feedback framework based on encoder-decoder is immutable. Inspired by the standardization scheme \cite{ref8}, this paper proposes a changeable-rate CSI feedback scheme, which is shown in Fig. \ref{fig4}, to improve the usability of DL-based CSI feedback architectures. This paper introduces a general feedback overhead control unit (FOCU) for autoencoder architectures to constrain the length of feedback codeword. Because it is not supported to change the length of vectors in the FC layer, after obtaining the $M\times1$ dimension-compressed codeword $\left[c_1,c_2,\ldots,c_M\right]$, FOCU discards the last part of the codeword directly. The remaining codeword of length $n\left(\le M\right)$, i.e., $\left[c_1,c_2,\ldots,c_n\right]$, is fed back to the BS side through the feedback link. Note that $n$ can be different during each transmission. Also, because the operation in the FC layer is length-fixed, decoder cannot immediately exploit the fed back codeword of length $n$. To address this issue, FOCU builds a placeholder (vector) of size $M\times1$ before decoder at the BS side. The feedback codeword is the first $n$ elements of the placeholder. The last $M-n$ elements of the placeholder are zeros, i.e., the codeword becomes $[c_1,c_2,\ldots,c_n,{\underbrace{0,\ldots,0}_{(M-n) \text{ zeros}}}]$. Finally, decoder reconstructs the CSI matrix from the zero-padded codeword.

FOCU makes changeable feedback overhead possible. Properly trained autoencoder with FOCU supports at most $M+1$ feedback overhead options. To train the neural network with FOCU, we first need to specify the length of feedback codeword $n\in\left\{0,1,\ldots,M\right\}$ for each training sample. During the training process, FOCU discards the last $M-n$ elements of the codeword at the UE side. At the BS side, FOCU zero-pads $M-n$ zeros to the truncated codeword to guarantee that the length of the codeword $M$ is fixed for the decoder. MSE is used as the loss function, which is formulated as follows
  \begin{equation}
    \begin{split}
      \text{LOSS}=\frac{1}{NM}\sum_{k=1}^N\sum_{n=0}^M\lambda_n\Vert\mathbf{H}_d^k-f_\text{de}(f_\text{de-quan}(f_\text{quan}(f_\text{en}(\mathbf{H}_d^k,n))))\Vert_2^2,\label{eq_10} 
    \end{split}
  \end{equation}
where $N$ is the size of training dataset, and $k$ is the index of training data. $\lambda_n$ is the weight coefficient of feedback codeword with the length of $M-n$. In this work, we set $n$ as a random variable that is uniformly distributed over $\left\{0,1,\ldots,M\right\}$. Therefore, we set $\lambda_n=1$ for all $n\in\left\{0,1,\ldots,M\right\}$. According to \eqref{eq_10}, the neural network with FOCU is trained with feedback codewords of different lengths which vary from $0$ to $M$. So FOCU can reutilize all of the neural network layers to extract the features and recover original CSI from the feedback codewords of different lengths, and it is compatible with any other CSI feedback framework. For traditional length-fixed CSI encoding-decoding networks, the loss function can also be formulated as \eqref{eq_10}, where $\lambda_n=1$ for $n=0$ and $\lambda_n=0$ otherwise.

FOCU can be applied to any traditional DL-based length-fixed CSI feedback network to obtain a variable length CSI encoding-decoding network, which improves the efficiency of CSI encoding-decoding systems. To further demonstrate the advantages of FOCU, we build two changeable-rate CSI feedback networks named CH-CsiNetPro and CH-DualNetSph by integrating FOCU into CsiNetPro and DualNetSph, respectively.

\subsection{CH-CsiNetPro and CH-DualNetSph}
FOCU, directly connected to CsiNetPro and DualNetSph, can achieve changeable overhead for CSI feedback system by deploy only one transmitter (encoder) and one receiver (decoder) at the UE side and the BS side, so that it can improve the storage and recognize efficiency. CsiNetPro and DualNetSph with changeable-rate module FOCU are named as CH-CsiNetPro and CH-DualNetSph, respectively. Specifically, the encoder of CH-CsiNetPro or CH-DualNetSph compresses CSI matrix into an $M$-element codeword $\mathbf{s}_M$. Then, FOCU keeps the first $n$ elements of $\mathbf{s}_M$ and discards the rest at the encoder. At the decoder, the truncated codeword is zero-padded into the vector of length $M$ by FOCU and denoted as ${\mathbf{s}_M}^\prime$. Finally, ${\mathbf{s}_M}^\prime$ is utilized to recover the downlink CSI matrix. $M$ is the maximum length of the feedback codeword supported by the neural network. In our design, the value of $M$ is equal to $512$ for CH-CsiNetPro and $256$ for CH-DualNetSph. Thus, they support any feedback overhead less than or equal to $512$ and $256$, respectively.

\begin{table}[]
  \centering
  \caption{Trainable Parameters of Changeable and Fixed Feedback Overhead Networks at the UE and the BS Side.}
	\label{table2}
  \setlength{\tabcolsep}{3.7mm}{
    \begin{tabular}{c|c|c|c}
      \hline
      Network       & UE          & BS          & Total       \\ \hline
      CsiNetPro     & 2,083,038   & 2,091,966   & 4,175,004   \\ \hline
      CH-CsiNetPro  & 1,059,174   & 1,060,646   & 2,119,820   \\ \hline
      Reduce by     & 49.152\%    & 49.300\%    & 49.226\%    \\ \hline
      DualNetSph    & 553,925     & 562,149     & 1,116,074   \\ \hline
      CH-DualNetSph & 271,505     & 272,993     & 544,498     \\ \hline
      Reduce by     & 50.985\%    & 51.438\%    & 51.213\%    \\ \hline
      \end{tabular}
  }
\end{table}

CH-CsiNetPro and CH-DualNetSph are trained in an end-to-end way using a large-scale dataset generated by COST 2100 \cite{ref28}. For each sample of training data, the length of the truncated feedback codeword $n$ is uniformly and randomly realized over the set $\{0,1,\ldots,M\}$.

All parameters of CH-CsiNetPro and CH-DualNetSph are shared for different feedback overhead. Therefore, the features, extracted by the neural network, are the same for the expression and reconstruction of any length of codeword. The number of trainable parameters of CH-CsiNetPro and CH-DualNetSph are equal to that of CsiNetPro with $M=512$ and the DualNetSph with $M=256$, i.e., $P_{\rm CH-CsiNetPro}=2,119,820$ and $P_{\rm CH-DualNetSph}=544,498$, respectively. $P_{{\rm CsiNetPro},M}$ and $P_{{\rm DualNetSph},M}$ are used to denote the number of trainable parameters of CsiNetPro and DualNetSph in feedback codeword length $M$, respectively. The sum of the number of trainable parameters of CsiNetPro and DualNetSph is calculated as $\sum_{n\in\mathcal{N}} P_{{\rm CsiNetPro},n}$ and $\sum_{n\in\mathcal{N}} P_{{\rm DualNetSph},n}$, respectively, where $M\in\mathcal{N}\subset\{0,1,\ldots,M\}$. Therefore, the numbers of the parameters of the proposed changeable-rate networks are reduced by the factor of $1-\frac{P_{\rm CH-CsiNetPro}}{\sum_{n\in\mathcal{N}} P_{{\rm CsiNetPro},n}}$ for CH-CsiNetPro and $1-\frac{P_{\rm CH-DualNetSph}}{\sum_{n\in\mathcal{N}} P_{{\rm DualNetSph},n}}$ for CH-DualNetSph.

Consider a typical setting of length-fixed CSI feedback network, we set $\mathcal{N}=\{32,64,128,256,\\512\}$ for CsiNetPro and $\mathcal{N}=\left\{16,32,64,128,256\right\}$ for DualNetSph. For the two networks, $\mathcal{N}$ is the set of supported feedback overhead options. The total number of trainable parameters of CsiNetPro is $\sum_{n\in\mathcal{N}}P_{{\rm CsiNetPro},n}=4,175,004$, and that number of DualNetSph is $\sum_{n\in\mathcal{N}}P_{{\rm DualNetSph},n}=1,116,074$. Table \ref{table2} shows the storage overhead at the UE and the BS sides using CH-CsiNetPro, CH-DualNetSph and CsiNetPro, DualNetSph with the typical settings of feedback overhead. The supported lengths of feedback codewords of CH-CsiNetPro are $1$ to $512$, and that of CH-DualNetSph are $1$ to $256$. Note that not only can CH-CsiNetPro and CH-DualNetSph significantly improve the storage efficiency at both the UE and the BS sides, but also support more feedback overhead options compared with the typical settings of CsiNetPro and DualNetSph implementations. CH-CsiNetPro can reduce the total storage space by $49\%$ compared with deploying multiple fixed-rate networks, one for each of the feedback overhead options. For CH-DualNetSph, the storage space is reduced by $51\%$.

Besides the number of trainable parameters, the number of FLOPs of a neural network is also an important metric, which measures the computation overhead of the neural network model. The difference between the changeable-rate network and fixed-rate network is that FOCU (implemented in the changeable-rate network) modifies FC layers. The number of FLOPs of the FC layer is calculated as
\begin{equation}
  {\rm FLOPs}=2\times I\times O, \label{eq_11} 
\end{equation}
where $I$ and $O$ are the sizes of input and output FC layer, respectively. Assume that the length of feedback codeword is $L(<M)$ during the test of the model. At the UE side, the number of FLOPs of the FC layer of length-fixed network is $2\times2N_t{\widetilde{N}}_s\times L$. For the changeable-rate network, only the first $L$ elements of the FC layer need to be calculated and fed back to the BS. Therefore, the number of FLOPs of the FC layer is also $4N_t{\widetilde{N}}_sL$. At the BS side, only the first $L$ elements of the FC layer is involved in the computation of the decoder of changeable-rate network. The number of FLOPs is $2\times L\times2N_t{\widetilde{N}}_s$, which is the same as that of length-fixed network. Therefore, during the inference of the model, at both the UE and BS sides, the changeable-rate network with the FOCU does not increase the number of FLOPs compared with the fixed-rate network.

\section{Quantization with Bound Constraints and Approximate Gradient}

\begin{figure}[htbp]
  \centerline{\includegraphics[scale=0.487]{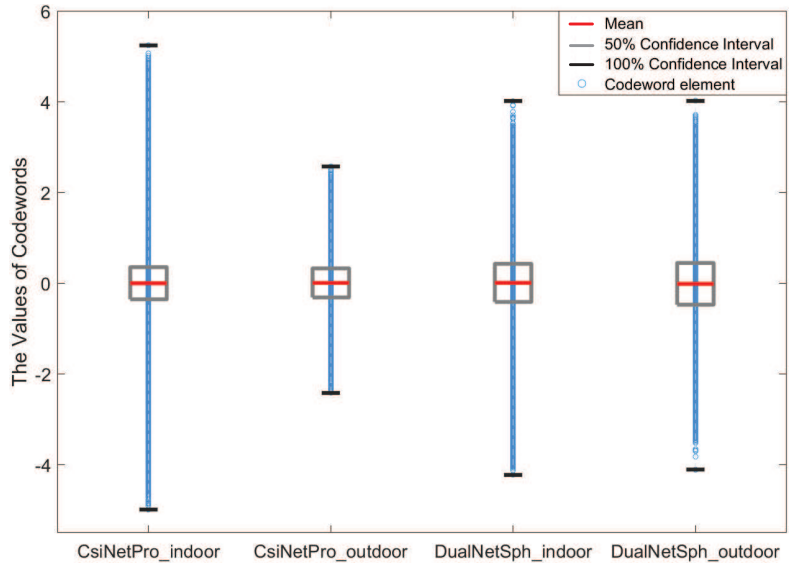}}
  \caption{The box chart of the values of codeword elements and their statistics.}
  \label{fig5} 
\end{figure}

Although the current DL-based CSI feedback frameworks can achieve outstanding performance, high accuracy of CSI reconstruction benefits from the typical computer number representation in DL-based algorithms \cite{ref18,ref19,ref20,ref23}, i.e., 32-bit floating-point number, which consumes a bulk of bandwidth resources. Compared with CsiNetPro, DualNetSph exploits MDPQ to quantize phase so that it is more efficient than CsiNetPro. However, the magnitude, which is compressed by the neural network, is still represented and fed back in 32-bit floating-point number and occupies much feedback link resource. Therefore, a general coding approach with less quantization-bit is required for DL-based CSI feedback system to improve the encoding efficiency of the neural network.

This section still takes CsiNetPro and DualNetSph as examples to analyze the distribution of encoded codeword to design the efficient quantization module. Then, we introduce the details of the proposed quantization module, i.e., pluggable quantization block (PQB).

The quantization operation divides the domain into a finite number of non-overlapping sub-intervals and the input of the quantizer fallen into each of the sub-interval is represented with a fixed value. Although the quantization operation introduces quantization error, which can be regarded as noise, it can greatly reduce the feedback overhead.

In \cite{ref29}, uniform quantization, the most basic and widely-used quantization approach, is utilized to represent compressed codewords. However, uniform quantization is deficient to represent non-uniformly distributed signals. Non-uniform quantizer, especially $\mu$-law compandor, is employed to quantize codewords in CSI feedback process to deal with the non-uniform signal quantization problem \cite{ref30}. The $\mu$-law transformation is defined as
\begin{equation}
  f\left(x\right)=\frac{\ln{(1+\mu x)}}{1+\mu},x\in[0,1], \label{eq_12} 
\end{equation}
where $x$ is the input signal and $\mu$ is the companding coefficient. Then, $f(x)$ is quantized uniformly.

The authors in \cite{deep_tasks} provide two end-to-end quantization schemes for DL-based tasks with bounded signal, i.e., passing gradient quantization and soft-to-hard quantization. The passing gradient quantization approach skips the back-propagation process of the quantization operation (this is equivalent to setting the gradient of the quantizer to constant one \cite{ref27}). The soft-to-hard quantization replaces the quantization function with an approximate differentiable function.

Fig. \ref{fig5} visualizes the distribution of the codewords and the statistics of well-trained CsiNetPro and DualNetSph in indoor and outdoor scenarios. Although most of codeword element values are concentrated around $0$, the maximum and minimum of these values are unbounded. To retain more information, many DL-based CSI feedback architectures, including CsiNetPro and DualNetSph, utilize the linear activation function to transform codeword values. This issue results in unbounded distribution of codewords. The above-mentioned quantization approaches, i.e., uniform, $\mu$-law quantization, passing gradient and soft-to-hard quantization, require the knowledge of the maximum and minimum values of codewords. However, these values are dynamic and hard to know in advance in the end-to-end neural networks training. A suboptimal solution is to train a network without quantization, then retrain the decoder with the quantized codewords. However, such a retraining process makes the global optimization of quantization operation and CSI reconstruction impossible, and consequently, it is hard for decoder to utilize all of the information carried by codewords. An end-to-end CSI feedback framework with quantization should be deployed. The values of the codewords must be bounded to enable the end-to-end training process.



\begin{figure*}[]
  \centerline{\includegraphics[scale=0.44]{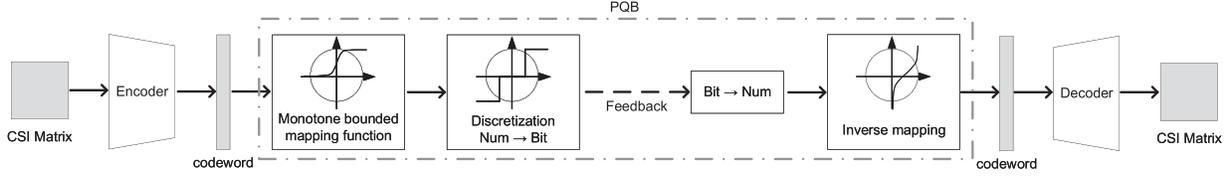}}
  \caption{The structure of PQB. Utilizing sigmoid function, it is possible to obtain non-linearly companded and bounded codewords simultaneously.}
  \label{fig7} 
\end{figure*}

\begin{figure}[htbp]
  \centerline{\includegraphics[scale=0.7]{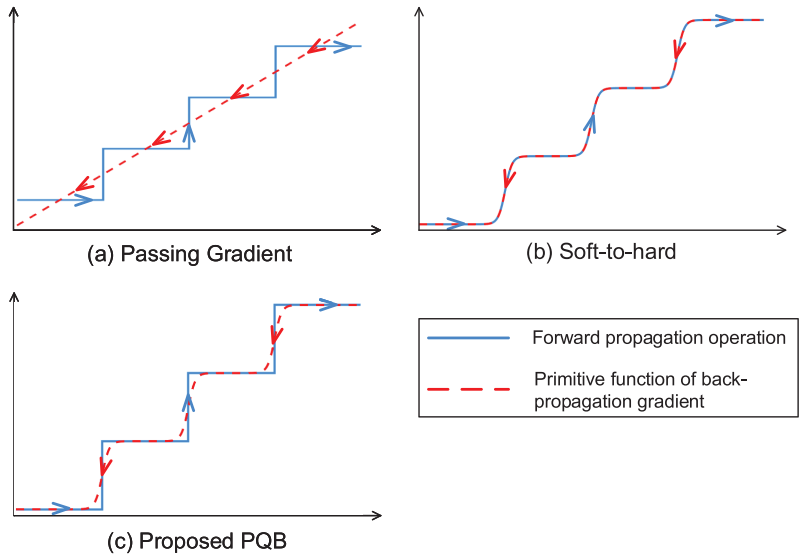}}
  \caption{The illustration of end-to-end quantization methods. The proposed PQB considers the behavior of quantizer in both forward and back-propagation processes.}
  \label{fig8} 
\end{figure}

Different from the solution of retraining the decoder, we consider a monotone bounded function defined over $\mathbb{R}$, i.e., sigmoid function formulated as $S(x)=\frac{1}{1+e^{-x}}$, for both non-linear transformation as well as bound constraints. As shown in Fig. \ref{fig7}, codeword $\mathbf{s}_M$ is firstly transformed by $S(x)$. Then, $S(\mathbf{s}_M)$ is quantized into a bit stream to feedback and de-quantized back into the approximated values $S({\hat{\mathbf{s}}}_M)$. Finally, inverse sigmoid function $S^{-1}(x)$ transforms $S({\hat{\mathbf{s}}}_M)$ into approximated truncated codeword ${\hat{\mathbf{s}}}_M$. Note that the described non-linear mapping approach is universal and it can work for any length of the codewords $\mathbf{s}_n$ as well. Therefore, DL-based CSI feedback frameworks with quantization can be trained in an end-to-end way.

The distribution of the elements of codewords cannot be obtained during the process of end-to-end training. Therefore, it is hard to design the quantizer according to the distribution before the training of neural networks. To make the neural network with randomly initialized parameters converges well at the early stage of training, for the interval $x\in[0,1]$, we consider the uniform quantizer, and its quantization and de-quantization functions are formulated as follows,

\begin{equation}
  f_\text{quan}(x)=\text{round}(2^b\times x-0.5), \label{eq_13} 
\end{equation}
\begin{equation}
  f_\text{de-quan}(x)=\frac{f_\text{quan}(x)+0.5}{2^b}, \label{eq_14} 
\end{equation}

\noindent where $\text{round}(\cdot)$ denotes the rounding function and $b$ is the number of quantization bits. The gradient of quantization operation does not exist everywhere, which makes the back-propagation training impossible.

To tackle this issue, as shown in Fig. \ref{fig8} (a) and (b), passing gradient quantization \cite{ref27,deep_tasks} skips the back-propagation process of the quantization operation, and it is equivalent to setting the gradient of the quantization operation to constant one. Soft-to-hard quantization \cite{deep_tasks} replaces the quantization function \eqref{eq_13} with a differentiable function formulated as follows
\begin{equation}
  \widetilde{f_{quan}}\left(x\right)=\sum_{i=1}^{2^b-1}{0.5(\tanh{\left(a\left(2^bx-i\right)\right)}+1)},x\in\left[0,1\right], \label{eq_15} 
\end{equation}
\noindent where $b$ is the number of quantization bits and $a$ is a parameter that controls the degree of approximation.

Different from the passing gradient and soft-to-hard quantization approaches, this paper proposes an approximate gradient for the back-propagation process of the quantization function $f_{quan}\left(x\right)$, which is formulated as
\begin{equation}
  \begin{aligned} 
  grad(x)=\left\{
    \begin{aligned}  
      \begin{array}{lr}
        \frac{1}{C\times d}\exp{\left(-\frac{1}{1-\left(\frac{M\left(x\right)}{d}\right)^2}\right)},M\left(x\right)\in\left(-d,d\right)\\
        \ \ \ \ \ \ \ \ \ \ \ \ \ \ \ 0\ \ \ \ \ \ \ \ \ \ \ ,\ \ \ \ \text{otherwise}
      \end{array}
    \end{aligned}
  \right.,x\in[0,1],
  \end{aligned}
  \label{eq_16} 
\end{equation}
where $M\left(x\right)=(x\text{ mod }{\frac{1}{2^b}})-\frac{1}{2^{b+1}}$, and $d\in(0,\frac{1}{2^{b+1}}]$ controls the degree of approximation, i.e., the degree of approximation to the Dirac's delta function. $C$ is the normalization factor. The approximate gradient of the quantizer $grad\left(x\right)$ exists everywhere. During the training process, the original gradient of quantization operation is replaced with $grad\left(x\right)$.

To explain the advantages of the proposed PQB, we have two remarks as follows:

{\bf \emph{Remark 1}:} Indeed, the proposed approximate gradient function is constructed upon the function $h(x)=\exp \left(-\frac{1}{1-x^2}\right), x\in(-1,1)$, which is a bump function. In other words, $h(x)$ is smoooth and compactly supported. Note that the Dirac's delta function can be defined as follows
\begin{align}
\delta(x)=\lim_{\epsilon\rightarrow 0^+} \epsilon^{-1}g\left(\frac{x}{\epsilon}\right),\label{eq:00}
\end{align}
where $g(x)$ is an absolutely integrable function of total integral $1$. For the case where $g(x)$ is a bump function, it is guaranteed that the limit in \eqref{eq:00} converges to Dirac's delta function almost everywhere\cite{WalterRudin0Functional}. Therefore, it is desirable to use a bump function, e.g., $h(x)$, as an approximation of Dirac's delta function.

\begin{figure}[htbp]
  \centerline{\includegraphics[scale=0.63]{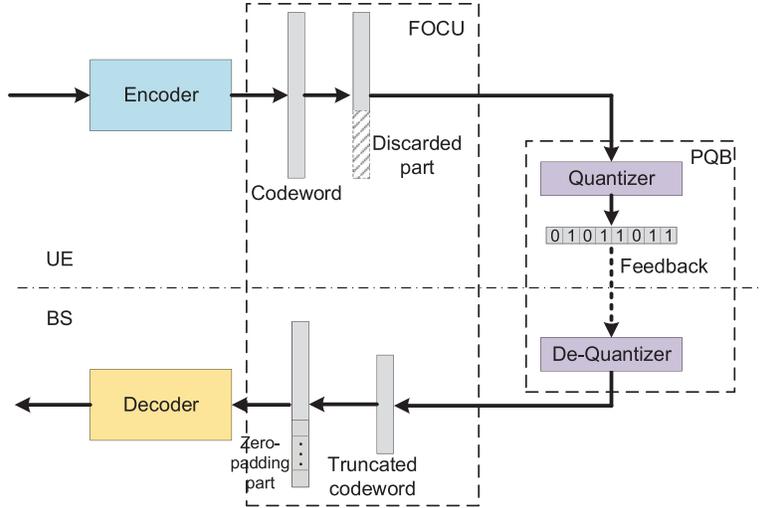}}
  \caption{Structure of proposed CH-CsiNetPro-PQB and CH-DualNetSph-PQB. FOCU and PQB can jointly optimize the efficiency of CSI feedback systems.}
  \label{fig8.5} 
\end{figure}

{\bf \emph{Remark 2}:} The passing gradient quantization skips the back-propagation process of $f_{quan}\left(x\right)$ directly and the soft-to-hard quantization replaces the forward propagation function $f_{quan}\left(x\right)$ with $\widetilde{f_{quan}}\left(x\right)$ to enable the training process. Due to the fact that the behavior of the quantizer is deterministic, it is desirable to take the process of forward and back-propagation of the quantization operation in the end-to-end training into consideration. As shown in Fig. \ref{fig8} (c), the proposed quantization scheme indeed takes the behavior of the quantizer into consideration. The quantizer uses $f_{quan}\left(x\right)$ in forward propagation. In the process of back-propagation, we use $grad\left(x\right)$ as an approximation to mimic the behavior of $f_{quan}\left(x\right)$.

In summary, the proposed PQB is a plug-in module for improving the encoding efficiency of DL-based CSI feedback frameworks. The bounded transformation and approximate gradient of the quantizer make it possible to train the neural networks with PQB in an end-to-end way. The behavior of the quantizer is completely considered and the global optimization of the CSI feedback networks can be achieved.

Since PQB is valid for truncated codeword ${\mathbf{s}_M}^\prime$ as well, PQB can be combined with changeable-rate CSI feedback frameworks proposed in Section \uppercase\expandafter{\romannumeral3} to jointly improve both the encoding and the storage efficiency of CSI feedback system. Specifically, codeword generated from the encoder is firstly truncated because of the feedback overhead constraint. Then, truncated codeword is quantized by PQB and zero-padded into the codeword of the original length. Finally, quantized and zero-padded codeword is utilized to recover CSI. As shown in Fig. \ref{fig8.5}, two joint efficiency-optimizing frameworks, CH-CsiNetPro-PQB and CH-DualNetSph-PQB, are proposed. They can be trained in an end-to-end way to achieve length-changeable CSI feedback and codewords quantization simultaneously. Compared with CsiNetPro and DualNetSph, the novelty of CH-CsiNetPro-PQB and CH-DualNetSph-PQB is that they can achieve bit-level variable length encoding and decoding for DL-based CSI feedback systems, and it provides a guideline for future research on DL-based changeable-rate CSI feedback.


\section{Performance Evaluation}

We first summarize the proposed methods as follows:

\begin{enumerate}[]
  \item[*] FOCU. Make the feedback overhead of DL-based CSI feedback network changeable. We apply FOCU to two DL-based fixed-rate CSI feedback networks CsiNetPro and DualNetSph, and propose two changeable-rate CSI feedback networks, named CH-CsiNetPro and CH-DualNetSph.
  \item[*] PQB. A quantizer that can make the DL-based CSI feedback network achieve an end-to-end quantization. PQB fully considers the behavior of the quantization operation in the forward and back-propagation process. We apply PQB to CsiNetPro and DualNetSph, and propose their counterparts which are with quantization operation, named CsiNetPro-PQB and DualNetSph-PQB.
  \item[*] FOCU+PQB. The combination of FOCU and PQB can make the DL-based CSI feedback networks achieve bit-level changeable rate CSI feedback. We apply FOCU+PQB to CsiNetPro and DualNetSph, and propose two networks named CH-CsiNetPro-PQB and CH-DualNetSph-PQB.
\end{enumerate} 

Then, this section introduces the experiment settings, including dataset generation, the training setting of neural networks and the evaluation method. Next, we evaluate the performance of proposed changeable-rate CSI feedback frameworks and analyze the mechanism behind FOCU. Finally, we evaluate the performance of proposed quantization module PQB and analyze the impact of PQB to the encoding efficiency of the neural networks.

\subsection{Dataset Description and Experiment Settings}
The widely-used channel model COST 2100 \cite{ref28} is adopted to generate the datasets of massive MIMO channels. Two scenarios are taken into consideration:
\begin{enumerate}[]
  \item Indoor picocellular scenario with downlink carrier frequency of $5.3$ GHz and uplink carrier frequency of $5.1$ GHz.
  \item Outdoor rural scenario with downlink carrier frequency of $300$ MHz and uplink carrier frequency of $260$ MHz.
\end{enumerate}

\begin{figure}[htbp]
  \centerline{\includegraphics[scale=0.4]{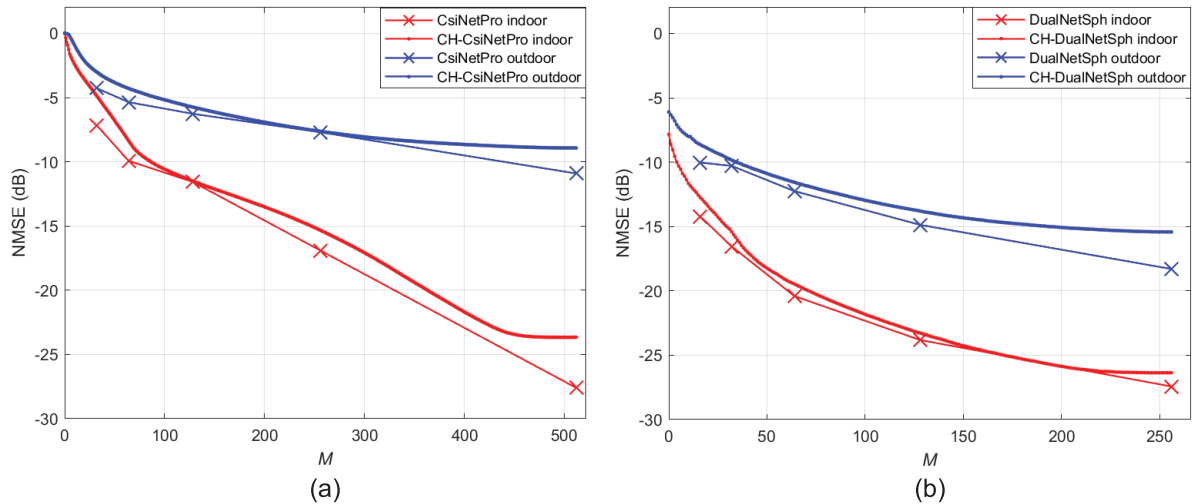}}
  \caption{Performance of proposed changeable-rate networks CH-CsiNetPro and CH-DualNetSph.}
  \label{fig9} 
\end{figure}

We place the BS at the center of a square area with the lengths of 20m and 400m for indoor and outdoor scenarios, respectively. The UE is uniformly randomly placed in the square area. This paper uses uniform linear array (ULA) with $N_t=32$ antennas at the BS side, and the spacing of antennas is set to half-wavelength. The bandwidth of downlink and uplink is both $20$ MHz for indoor and outdoor scenarios. The number of subcarriers is set to $N_s=1024$. After the transformation of the channel matrix from the spatial-frequency domain to the angular-delay domain using 2-D DFT, the CSI matrix becomes sparse. We retrain the first ${\widetilde{N}}_s=32$ rows of the channel matrix to further reduce the dimension. The rest of the parameters for dataset generation follows the default settings as described in \cite{ref28}.

The sizes of training and testing datasets are $100,000$ and $30,000$, respectively. The training dataset is used to update trainable parameters of the network and the testing dataset is used to evaluate the training results. The training and testing sets are disjoint. Similar to \cite{ref18}, matrices, as the input of the neural network, are normalized to the interval $\left[0,1\right]$ to facilitate the training process.

Because the proposed FOCU and PQB are both plug-in modules, they are compatible with other neural networks and the training settings of the neural networks should be the same as the original settings. For all the evaluated neural networks, we set the number of training epoch and batch size to $2000$ and $200$, respectively. The learning rate is fixed to $0.001$ and the loss function is MSE. The ADAM \cite{ref31} optimizer is used to update the trainable parameters.

The NMSE is used to evaluate the reconstruction accuracy of the downlink CSI matrix $\mathbf{H}_d$, NMSE is formulated as follows:

\begin{figure}[htbp]
  \centerline{\includegraphics[scale=0.4]{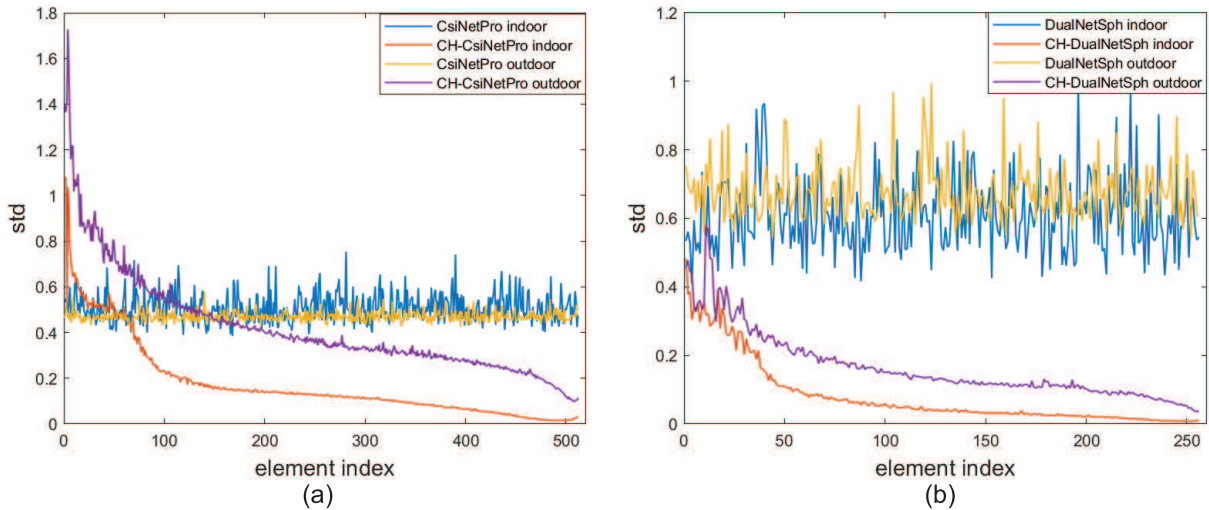}}
  \caption{The standard deviations comparison in the indoor environment and the outdoor environment.}
  \label{fig10} 
\end{figure}

\begin{equation}
  \text{NMSE}=\frac{1}{N}\sum_{k=1}^N\frac{\Vert{\mathbf{H}_d^k-\hat{\mathbf{H}}_d^k}\Vert_2^2}{\Vert{\mathbf{H}_d^k}\Vert_2^2}, \label{eq_17} 
\end{equation}
\noindent where $N$ is the size of training dataset.

\subsection{Evaluation of Changeable-Rate Feedback Networks}
\subsubsection{Performance of the Proposed CH-CsiNetPro and CH-DualNetSph}

Fig. \ref{fig9} (a) and (b) compare the NMSE performance in dB scale of the proposed changeable-rate CSI feedback networks CH-CsiNetPro and CH-DualNetSph with their fixed feedback overhead counterparts CsiNetPro and DualNetSph. The maximum feedback overhead is set to $M=512$ for CH-CsiNetPro and $M=256$ for CH-DualNetSph. The performances of CH-CsiNetPro and CH-DualNetSph with various feedback overhead values $\{0,1,\ldots,M\}$ are evaluated. For comparison, we train and test the fixed feedback overhead networks under $M=32$, $64$, $128$, $256$, $512$ for CsiNetPro and $M=16$, $32$, $64$, $128$, $256$ for DualNetSph.

From Fig. \ref{fig9}, it is obvious that CH-CsiNetPro and CH-DualNetSph use only one network to achieve nearly the same performance compared with the fixed feedback overhead networks in both indoor environment and outdoor environment. Specifically, for indoor scenario, when the feedback codeword is of medium length, the performance of CH-CsiNetPro is slightly worse than that of CsiNetPro by $0.86$ dB in average. When the length of feedback codeword is large or small, the performance degrades more, i.e., by $3.94$ dB for codeword of length $512$ and by $2.31$ dB for codeword of length $32$. For outdoor scenario, CH-CsiNetPro has a performance degradation of $0.95$ dB in average compared with CsiNetPro over all feedback overhead options. Similar result is also observed for CH-DualNetSph. This shows that the proposed FOCU has robust performance for changeable-rate feedback networks.

\subsubsection{Network Robustness Analyze}
As discussed in 1) of this subsection, CH-CsiNetPro and CH-DualNetSph, which only use one neural network respectively, can achieve almost the same accuracy compared with fixed feedback overhead networks CsiNetPro and DualNetSph. In this part, the accuracy-ensuring and efficiency-improving mechanisms behind CH-CsiNetPro and CH-DualNetSph are analyzed. The CH-CsiNetPro and CH-DualNetSph adopt linear activation for the generation of codewords. This work firstly calculates the mean values of codewords generated by CH-CsiNetPro and CH-DualNetSph besides their fixed feedback overhead versions with $M=512$ and $256$, respectively. Statistical results show that the mean values of codewords in the above networks are near zero. Next, the standard deviation of the codeword values is shown in Fig. \ref{fig10} (a) and (b). The standard deviation (SD) can reflect the degree of dispersion of the codeword.

In Fig. \ref{fig10} (a), it is observed that in CH-CsiNetPro, the SDs of the values of the codewords with smaller indices is in general greater than that of the codewords with larger indices. But the SDs of the values of CsiNetPro codeword are much more concentrated and are independent of their index. The CH-DualNetSph in Fig. \ref{fig10} (b) has the similar observation. Considering the mechanism of FOCU, it always discards the last $M-n$ elements of the codeword under the constraint of the feedback overhead $n$ and applies zero-padding operation. Thus, changeable-rate networks tend to retain more information when the constraint of the feedback overhead $n$ is smaller.

Fig. \ref{fig9} (b) shows the NMSE of CH-DualNetSph. This work takes the indoor environment as an example, when the length of feedback codeword decreases from $256$ to $40$, the degradation of NMSE is $9.38$ dB. As the length of feedback codeword continues to decrease to $0$, the reduction of NMSE is $9.14$ dB. The performance results further show that the robustness of proposed changeable-rate feedback networks.

\subsection{Quantization Module Evaluation}
This subsection compares the proposed end-to-end quantization neural networks using PQB with the following quantization approaches:

\begin{table*}[]
  \centering
  \caption{The NMSE ($dB$) performance of PQB using CsiNetPro and DualNetSph. \protect\\Methods being Compared: Without Quantization (ORI); $\mu$-Law Quantization ($\mu$Q); \protect\\Passing Gradient Quantization (PG); Soft-to-Hard Quantization (S2H).}
	\label{table4}
  \resizebox{\textwidth}{!}
  {
    \begin{tabular}{cccccccccccccccccccccc}
      \hline
      \multicolumn{22}{c}{CsiNetPro}                                                                                                                                                                                                                                                                                                                                                                                                                                                                                                          \\ \hline
      \multicolumn{1}{c|}{\multirow{7}{*}{indoor}}  & \multicolumn{1}{c|}{$M$}    & \multicolumn{4}{c|}{32}                                                                    & \multicolumn{4}{c|}{64}                                                                    & \multicolumn{4}{c|}{128}                                                                   & \multicolumn{4}{c|}{256}                                                                   & \multicolumn{4}{c}{512}                                               \\ \cline{2-22} 
      \multicolumn{1}{c|}{}                         & \multicolumn{1}{c|}{$b$}    & 2               & 3               & 4               & \multicolumn{1}{c|}{5}               & 2               & 3               & 4               & \multicolumn{1}{c|}{5}               & 2               & 3               & 4               & \multicolumn{1}{c|}{5}               & 2               & 3               & 4               & \multicolumn{1}{c|}{5}               & 2               & 3               & 4               & 5               \\ \cline{2-22} 
      \multicolumn{1}{c|}{}                         & \multicolumn{1}{c|}{ORI}    & \multicolumn{4}{c|}{-7.16}                                                                 & \multicolumn{4}{c|}{-9.93}                                                                 & \multicolumn{4}{c|}{-11.56}                                                                & \multicolumn{4}{c|}{-16.91}                                                                & \multicolumn{4}{c}{-27.60}                                            \\ \cline{2-22} 
      \multicolumn{1}{c|}{}                         & \multicolumn{1}{c|}{$\mu$Q} & -4.16           & -5.11           & -6.28           & \multicolumn{1}{c|}{-6.84}           & -5.54           & -6.60           & -7.61           & \multicolumn{1}{c|}{-9.35}           & -7.01           & -7.96           & -9.95           & \multicolumn{1}{c|}{-11.12}          & -9.80           & -10.28          & -13.63          & \multicolumn{1}{c|}{-15.39}          & -12.15          & -12.68          & -15.80          & -19.85          \\ \cline{2-22} 
      \multicolumn{1}{c|}{}                         & \multicolumn{1}{c|}{PG}     & \textbf{-5.68}  & -6.56           & -6.75           & \multicolumn{1}{c|}{-6.91}           & -7.71           & -9.06           & -9.53           & \multicolumn{1}{c|}{-9.89}           & -8.72           & -9.92           & -10.77          & \multicolumn{1}{c|}{-11.27}          & -11.90          & -13.12          & -13.65          & \multicolumn{1}{c|}{-16.47}          & -13.78          & -16.05          & -18.05          & -22.80          \\ \cline{2-22} 
      \multicolumn{1}{c|}{}                         & \multicolumn{1}{c|}{S2H}    & {\color{red}+3.97}           & -2.65           & -3.96           & \multicolumn{1}{c|}{-5.79}           & {\color{red}+6.33}           & -1.75           & -6.27           & \multicolumn{1}{c|}{-7.97}           & {\color{red}+5.90}           & -6.03           & -7.21           & \multicolumn{1}{c|}{-11.86}          & {\color{red}+5.82}           & -8.74           & -10.80          & \multicolumn{1}{c|}{-11.51}          & -2.52           & -11.09          & -18.43          & -22.27          \\ \cline{2-22} 
      \multicolumn{1}{c|}{}                         & \multicolumn{1}{c|}{PQB}    & \textbf{-5.68}  & \textbf{-6.71}  & \textbf{-7.02}  & \multicolumn{1}{c|}{\textbf{-7.33}}  & \textbf{-7.95}  & \textbf{-9.36}  & \textbf{-9.91}  & \multicolumn{1}{c|}{\textbf{-10.04}} & \textbf{-10.22} & \textbf{-10.24} & \textbf{-11.18} & \multicolumn{1}{c|}{\textbf{-12.06}} & \textbf{-12.22} & \textbf{-13.62} & \textbf{-14.84} & \multicolumn{1}{c|}{\textbf{-16.72}} & \textbf{-14.62} & \textbf{-17.92} & \textbf{-20.52} & \textbf{-23.89} \\ \hline
                                                    &                             &                 &                 &                 &                                      &                 &                 &                 &                                      &                 &                 &                 &                                      &                 &                 &                 &                                      &                 &                 &                 &                 \\ \hline
      \multicolumn{1}{c|}{\multirow{7}{*}{outdoor}} & \multicolumn{1}{c|}{$M$}    & \multicolumn{4}{c|}{32}                                                                    & \multicolumn{4}{c|}{64}                                                                    & \multicolumn{4}{c|}{128}                                                                   & \multicolumn{4}{c|}{256}                                                                   & \multicolumn{4}{c}{512}                                               \\ \cline{2-22} 
      \multicolumn{1}{c|}{}                         & \multicolumn{1}{c|}{$b$}    & 2               & 3               & 4               & \multicolumn{1}{c|}{5}               & 2               & 3               & 4               & \multicolumn{1}{c|}{5}               & 2               & 3               & 4               & \multicolumn{1}{c|}{5}               & 2               & 3               & 4               & \multicolumn{1}{c|}{5}               & 2               & 3               & 4               & 5               \\ \cline{2-22} 
      \multicolumn{1}{c|}{}                         & \multicolumn{1}{c|}{ORI}    & \multicolumn{4}{c|}{-4.27}                                                                 & \multicolumn{4}{c|}{-5.36}                                                                 & \multicolumn{4}{c|}{-6.26}                                                                 & \multicolumn{4}{c|}{-7.70}                                                                 & \multicolumn{4}{c}{-10.91}                                            \\ \cline{2-22} 
      \multicolumn{1}{c|}{}                         & \multicolumn{1}{c|}{$\mu$Q} & -2.62           & -3.17           & -3.83           & \multicolumn{1}{c|}{-4.21}           & -3.77           & -4.19           & -4.89           & \multicolumn{1}{c|}{-5.16}           & -4.69           & -5.30           & -5.95           & \multicolumn{1}{c|}{-6.20}           & -6.00           & -6.63           & -7.40           & \multicolumn{1}{c|}{-7.75}           & -7.82           & -8.55           & -10.02          & -10.39          \\ \cline{2-22} 
      \multicolumn{1}{c|}{}                         & \multicolumn{1}{c|}{PG}     & -3.52           & -3.76           & -3.95           & \multicolumn{1}{c|}{-4.28}           & -4.31           & -4.75           & -4.93           & \multicolumn{1}{c|}{-5.00}           & -5.02           & -5.37           & -6.10           & \multicolumn{1}{c|}{-6.14}           & -6.57           & \textbf{-7.58}  & -7.70           & \multicolumn{1}{c|}{-7.87}           & -8.72           & -9.64           & -10.13          & -10.38          \\ \cline{2-22} 
      \multicolumn{1}{c|}{}                         & \multicolumn{1}{c|}{S2H}    & -2.68           & -3.98           & -4.22           & \multicolumn{1}{c|}{-4.37}           & -3.61           & -4.70           & -5.10           & \multicolumn{1}{c|}{-5.17}           & -4.70           & -5.52           & -6.04           & \multicolumn{1}{c|}{-6.13}           & -6.08           & -7.37           & -7.51           & \multicolumn{1}{c|}{-7.45}           & -7.72           & -9.66           & -9.89           & -10.54          \\ \cline{2-22} 
      \multicolumn{1}{c|}{}                         & \multicolumn{1}{c|}{PQB}    & \textbf{-3.58}  & \textbf{-4.11}  & \textbf{-4.26}  & \multicolumn{1}{c|}{\textbf{-4.49}}  & \textbf{-4.41}  & \textbf{-4.88}  & \textbf{-5.10}  & \multicolumn{1}{c|}{\textbf{-5.19}}  & \textbf{-5.38}  & \textbf{-5.88}  & \textbf{-6.21}  & \multicolumn{1}{c|}{\textbf{-6.21}}  & \textbf{-6.78}  & \textbf{-7.58}  & \textbf{-7.84}  & \multicolumn{1}{c|}{\textbf{-7.91}}  & \textbf{-8.80}  & \textbf{-9.89}  & \textbf{-10.40} & \textbf{-10.77} \\ \hline
                                                    &                             &                 &                 &                 &                                      &                 &                 &                 &                                      &                 &                 &                 &                                      &                 &                 &                 &                                      &                 &                 &                 &                 \\ \hline
      \multicolumn{22}{c}{DualNetSph}                                                                                                                                                                                                                                                                                                                                                                                                                                                                                                         \\ \hline
      \multicolumn{1}{c|}{\multirow{7}{*}{indoor}}  & \multicolumn{1}{c|}{$M$}    & \multicolumn{4}{c|}{16}                                                                    & \multicolumn{4}{c|}{32}                                                                    & \multicolumn{4}{c|}{64}                                                                    & \multicolumn{4}{c|}{128}                                                                   & \multicolumn{4}{c}{256}                                               \\ \cline{2-22} 
      \multicolumn{1}{c|}{}                         & \multicolumn{1}{c|}{$b$}    & 2               & 3               & 4               & \multicolumn{1}{c|}{5}               & 2               & 3               & 4               & \multicolumn{1}{c|}{5}               & 2               & 3               & 4               & \multicolumn{1}{c|}{5}               & 2               & 3               & 4               & \multicolumn{1}{c|}{5}               & 2               & 3               & 4               & 5               \\ \cline{2-22} 
      \multicolumn{1}{c|}{}                         & \multicolumn{1}{c|}{ORI}    & \multicolumn{4}{c|}{-14.22}                                                                & \multicolumn{4}{c|}{-16.55}                                                                & \multicolumn{4}{c|}{-20.41}                                                                & \multicolumn{4}{c|}{-23.82}                                                                & \multicolumn{4}{c}{-27.43}                                            \\ \cline{2-22} 
      \multicolumn{1}{c|}{}                         & \multicolumn{1}{c|}{$\mu$Q} & -11.07          & -12.06          & -12.97          & \multicolumn{1}{c|}{-13.68}          & -12.42          & -13.18          & -14.36          & \multicolumn{1}{c|}{-15.64}          & -13.94          & -14.39          & -16.99          & \multicolumn{1}{c|}{-19.05}          & -15.32          & -15.95          & -19.10          & \multicolumn{1}{c|}{-21.62}          & -16.84          & -17.62          & -21.22          & -24.59          \\ \cline{2-22} 
      \multicolumn{1}{c|}{}                         & \multicolumn{1}{c|}{PG}     & -12.56          & -13.36          & -13.73          & \multicolumn{1}{c|}{-13.91}          & -14.24          & -15.52          & -16.05          & \multicolumn{1}{c|}{-16.35}          & -16.32          & -18.10          & -19.24          & \multicolumn{1}{c|}{-19.56}          & -18.28          & -20.47          & -21.89          & \multicolumn{1}{c|}{-22.45}          & -20.44          & -23.00          & -24.48          & -25.74          \\ \cline{2-22} 
      \multicolumn{1}{c|}{}                         & \multicolumn{1}{c|}{S2H}    & {\color{red}+10.23}          & {\color{red}+5.47}           & -12.97          & \multicolumn{1}{c|}{-14.00}          & {\color{red}+3.04}           & {\color{red}+3.93}           & -14.09          & \multicolumn{1}{c|}{-15.26}          & {\color{red}+5.94}           & {\color{red}+4.59}           & -18.70          & \multicolumn{1}{c|}{-19.52}          & {\color{red}+5.54}           & {\color{red}+3.87}           & -20.28          & \multicolumn{1}{c|}{-22.88}          & {\color{red}+4.55}           & {\color{red}+4.09}           & -24.54          & -25.71          \\ \cline{2-22} 
      \multicolumn{1}{c|}{}                         & \multicolumn{1}{c|}{PQB}    & \textbf{-12.59} & \textbf{-13.54} & \textbf{-13.91} & \multicolumn{1}{c|}{\textbf{-14.21}} & \textbf{-14.30} & \textbf{-15.60} & \textbf{-16.29} & \multicolumn{1}{c|}{\textbf{-16.73}} & \textbf{-16.34} & \textbf{-18.30} & \textbf{-19.58} & \multicolumn{1}{c|}{\textbf{-20.09}} & \textbf{-18.30} & \textbf{-20.81} & \textbf{-22.43} & \multicolumn{1}{c|}{\textbf{-23.41}} & \textbf{-20.50} & \textbf{-23.71} & \textbf{-26.38} & \textbf{-28.72} \\ \hline
                                                    &                             &                 &                 &                 &                                      &                 &                 &                 &                                      &                 &                 &                 &                                      &                 &                 &                 &                                      &                 &                 &                 &                 \\ \hline
      \multicolumn{1}{c|}{\multirow{7}{*}{outdoor}} & \multicolumn{1}{c|}{$M$}    & \multicolumn{4}{c|}{16}                                                                    & \multicolumn{4}{c|}{32}                                                                    & \multicolumn{4}{c|}{64}                                                                    & \multicolumn{4}{c|}{128}                                                                   & \multicolumn{4}{c}{256}                                               \\ \cline{2-22} 
      \multicolumn{1}{c|}{}                         & \multicolumn{1}{c|}{$b$}    & 2               & 3               & 4               & \multicolumn{1}{c|}{5}               & 2               & 3               & 4               & \multicolumn{1}{c|}{5}               & 2               & 3               & 4               & \multicolumn{1}{c|}{5}               & 2               & 3               & 4               & \multicolumn{1}{c|}{5}               & 2               & 3               & 4               & 5               \\ \cline{2-22} 
      \multicolumn{1}{c|}{}                         & \multicolumn{1}{c|}{ORI}    & \multicolumn{4}{c|}{-10.03}                                                                & \multicolumn{4}{c|}{-10.29}                                                                & \multicolumn{4}{c|}{-12.25}                                                                & \multicolumn{4}{c|}{-14.87}                                                                & \multicolumn{4}{c}{-18.30}                                            \\ \cline{2-22} 
      \multicolumn{1}{c|}{}                         & \multicolumn{1}{c|}{$\mu$Q} & -8.28           & -8.74           & -9.39           & \multicolumn{1}{c|}{-9.58}           & -8.56           & -8.81           & -9.82           & \multicolumn{1}{c|}{-10.25}          & -9.61           & -10.09          & -11.35          & \multicolumn{1}{c|}{-12.10}          & -10.93          & -11.62          & -13.34          & \multicolumn{1}{c|}{-14.28}          & -12.75          & -13.56          & -15.76          & -17.51          \\ \cline{2-22} 
      \multicolumn{1}{c|}{}                         & \multicolumn{1}{c|}{PG}     & -9.18           & -9.46           & -9.77           & \multicolumn{1}{c|}{-9.93}           & -9.89           & -10.16          & -10.47          & \multicolumn{1}{c|}{-10.49}          & -11.15          & -11.45          & -11.80          & \multicolumn{1}{c|}{-11.83}          & -12.58          & -13.32          & -13.67          & \multicolumn{1}{c|}{-13.89}          & -14.74          & -15.88          & -16.26          & -16.74          \\ \cline{2-22} 
      \multicolumn{1}{c|}{}                         & \multicolumn{1}{c|}{S2H}    & {\color{red}+7.78}           & {\color{red}+8.10}           & -9.33           & \multicolumn{1}{c|}{-9.84}           & {\color{red}+11.01}          & {\color{red}+9.45}           & -10.59          & \multicolumn{1}{c|}{-10.69}          & {\color{red}+13.68}          & {\color{red}+6.28}           & -11.93          & \multicolumn{1}{c|}{-12.00}          & {\color{red}+7.76}           & {\color{red}+4.74}           & -13.76          & \multicolumn{1}{c|}{-14.45}          & {\color{red}+4.52}           & -16.02          & -16.99          & -17.50          \\ \cline{2-22} 
      \multicolumn{1}{c|}{}                         & \multicolumn{1}{c|}{PQB}    & \textbf{-9.24}  & \textbf{-9.56}  & \textbf{-9.90}  & \multicolumn{1}{c|}{\textbf{-10.14}} & \textbf{-9.97}  & \textbf{-10.53} & \textbf{-10.79} & \multicolumn{1}{c|}{\textbf{-10.86}} & \textbf{-11.23} & \textbf{-11.77} & \textbf{-12.24} & \multicolumn{1}{c|}{\textbf{-12.38}} & \textbf{-12.66} & \textbf{-13.65} & \textbf{-14.32} & \multicolumn{1}{c|}{\textbf{-14.75}} & \textbf{-14.80} & \textbf{-16.25} & \textbf{-17.47} & \textbf{-17.93} \\ \hline
      \end{tabular}
  }
\end{table*}

\begin{itemize}
  \item[*] $\mu$-law quantization \cite{ref30} ($\mu$Q). The training process of neural network is divided into two phases. The neural network is firstly trained without quantization. Then, the decoder is retrained with the quantized codewords following $\mu$-law.
  \item[*]	Passing gradient quantization \cite{ref27,deep_tasks}. The neural network with quantization is trained in an end-to-end way. The back-propagation of the quantizer is skipped, which is equivalent to setting the gradient of the quantization function to constant one.
  \item[*]	Soft-to-hard quantization \cite{deep_tasks}. The quantization function is replaced with an approximate differentiable function. The neural network with the approximate quantization function is trained in an end-to-end way.
\end{itemize}

We use CsiNetPro and DualNetSph to show the performance of the proposed quantization approaches. The number of pretraining epochs for neural networks without quantization is still $2000$. Extra $500$ training epochs are performed for the decoder to optimize the network with quantization operation. In this experiment, we set $a=8$ in \eqref{eq_15} and $d=0.5$ in \eqref{eq_16}. We compare the settings with different numbers of quantization bits $b=2,3,4,5$.

Table \ref{table4} shows the NMSE performance of proposed quantization module PQB using CsiNetPro and DualNetSph. The end-to-end quantization networks with PQB outperform other approaches by a margin. When the number of quantization bits is $5$, CsiNetPro and DualNetSph with PQB can achieve nearly the same performance as that without quantization. PQB significantly improves the encoding efficiency. Compared with $32$-bit float-pointing expression of codewords, PQB saves $84.4\%$ bit-width with the same accuracy. Moreover, when the number of bits decreases, PQB can ensure a robust performance even with a small number of quantization bits. For example, in indoor environment, when the number of quantization bits of DualNetSph decreases from $5$ to $2$, the performance of $\mu$Q degrades rapidly. The network with 2-bit soft-to-hard quantization turns out to be hard to converge (marked in red). PQB also outperforms passing gradient quantization. Therefore, PQB can provide better trade-off between the performance and efficiency. From $4$-bit encoding to $2$-bit encoding, the NMSE performance of PQB is only worsened by about $20\%$ in dB scale. According to this, we further expect that PQB can still have a relatively robust performance under even tighter bit-width limitation, e.g., $1$-bit encoding.

\begin{table*}[]
  \caption{The Comparison of Empirical Entropy ($bits/element$) of Different Quantization Methods. \protect\\The Numbers of Quantization Bits are $3$ and $4$. The Lengths of Codewords are $64$, $256$ for CsiNetPro and $32$, $128$ for DualNetSph.}
	\label{table5}
  \resizebox{\textwidth}{!}
  {
    \begin{tabular}{cccccclcccccc}
      \cline{1-6} \cline{8-13}
      \multicolumn{6}{c}{CsiNetPro}                                                                                                                                    &  & \multicolumn{6}{c}{DualNetSph}                                                                                                                                   \\ \cline{1-6} \cline{8-13} 
      \multicolumn{1}{c|}{\multirow{6}{*}{indoor}}  & \multicolumn{1}{c|}{$M$}    & \multicolumn{2}{c|}{64}                            & \multicolumn{2}{c}{256}       &  & \multicolumn{1}{c|}{\multirow{6}{*}{indoor}}  & \multicolumn{1}{c|}{$M$}    & \multicolumn{2}{c|}{32}                            & \multicolumn{2}{c}{128}       \\ \cline{2-6} \cline{9-13} 
      \multicolumn{1}{c|}{}                         & \multicolumn{1}{c|}{$b$}    & 3             & \multicolumn{1}{c|}{4}             & 3             & 4             &  & \multicolumn{1}{c|}{}                         & \multicolumn{1}{c|}{$b$}    & 3             & \multicolumn{1}{c|}{4}             & 3             & 4             \\ \cline{2-6} \cline{9-13} 
      \multicolumn{1}{c|}{}                         & \multicolumn{1}{c|}{$\mu$Q} & 2.61          & \multicolumn{1}{c|}{3.44}          & 2.53          & 3.43          &  & \multicolumn{1}{c|}{}                         & \multicolumn{1}{c|}{$\mu$Q} & 2.60          & \multicolumn{1}{c|}{3.46}          & 2.54          & 3.45          \\ \cline{2-6} \cline{9-13} 
      \multicolumn{1}{c|}{}                         & \multicolumn{1}{c|}{PG}     & 2.05          & \multicolumn{1}{c|}{2.88}          & 2.09          & 2.97          &  & \multicolumn{1}{c|}{}                         & \multicolumn{1}{c|}{PG}     & 2.02          & \multicolumn{1}{c|}{2.76}          & 1.94          & 2.66          \\ \cline{2-6} \cline{9-13} 
      \multicolumn{1}{c|}{}                         & \multicolumn{1}{c|}{S2H}    & 2.21          & \multicolumn{1}{c|}{3.00}          & 2.26          & 2.94          &  & \multicolumn{1}{c|}{}                         & \multicolumn{1}{c|}{S2H}    & 1.98          & \multicolumn{1}{c|}{2.90}          & 1.98          & 2.83          \\ \cline{2-6} \cline{9-13} 
      \multicolumn{1}{c|}{}                         & \multicolumn{1}{c|}{PQB}    & 2.24          & \multicolumn{1}{c|}{3.11}          & 2.27          & 3.14          &  & \multicolumn{1}{c|}{}                         & \multicolumn{1}{c|}{PQB}    & 2.25          & \multicolumn{1}{c|}{3.11}          & 2.24          & 3.09          \\ \cline{1-6} \cline{8-13} 
      \multicolumn{1}{c|}{\multirow{6}{*}{outdoor}} & \multicolumn{1}{c|}{$M$}    & \multicolumn{2}{c|}{64}                            & \multicolumn{2}{c}{256}       &  & \multicolumn{1}{c|}{\multirow{6}{*}{outdoor}} & \multicolumn{1}{c|}{$M$}    & \multicolumn{2}{c|}{32}                            & \multicolumn{2}{c}{128}       \\ \cline{2-6} \cline{9-13} 
      \multicolumn{1}{c|}{}                         & \multicolumn{1}{c|}{$b$}    & 3             & \multicolumn{1}{c|}{4}             & 3             & 4             &  & \multicolumn{1}{c|}{}                         & \multicolumn{1}{c|}{$b$}    & 3             & \multicolumn{1}{c|}{4}             & 3             & 4             \\ \cline{2-6} \cline{9-13} 
      \multicolumn{1}{c|}{}                         & \multicolumn{1}{c|}{$\mu$Q} & 2.58          & \multicolumn{1}{c|}{3.45}          & 2.59          & 3.42          &  & \multicolumn{1}{c|}{}                         & \multicolumn{1}{c|}{$\mu$Q} & 2.61          & \multicolumn{1}{c|}{3.45}          & 2.56          & 3.42          \\ \cline{2-6} \cline{9-13} 
      \multicolumn{1}{c|}{}                         & \multicolumn{1}{c|}{PG}     & 2.05          & \multicolumn{1}{c|}{2.71}          & 1.96          & 2.74          &  & \multicolumn{1}{c|}{}                         & \multicolumn{1}{c|}{PG}     & 2.06          & \multicolumn{1}{c|}{2.76}          & 1.99          & 2.67          \\ \cline{2-6} \cline{9-13} 
      \multicolumn{1}{c|}{}                         & \multicolumn{1}{c|}{S2H}    & 2.03          & \multicolumn{1}{c|}{2.84}          & 2.07          & 3.07          &  & \multicolumn{1}{c|}{}                         & \multicolumn{1}{c|}{S2H}    & 1.90          & \multicolumn{1}{c|}{2.68}          & 2.01          & 2.67          \\ \cline{2-6} \cline{9-13} 
      \multicolumn{1}{c|}{}                         & \multicolumn{1}{c|}{PQB}    & 2.25          & \multicolumn{1}{c|}{3.12}          & 2.25          & 3.13          &  & \multicolumn{1}{c|}{}                         & \multicolumn{1}{c|}{PQB}    & 2.27          & \multicolumn{1}{c|}{3.12}          & 2.26          & 3.12          \\ \cline{1-6} \cline{8-13} 
      \end{tabular}
  }
\end{table*}

\begin{figure}[htbp]
  \centerline{\includegraphics[scale=0.6]{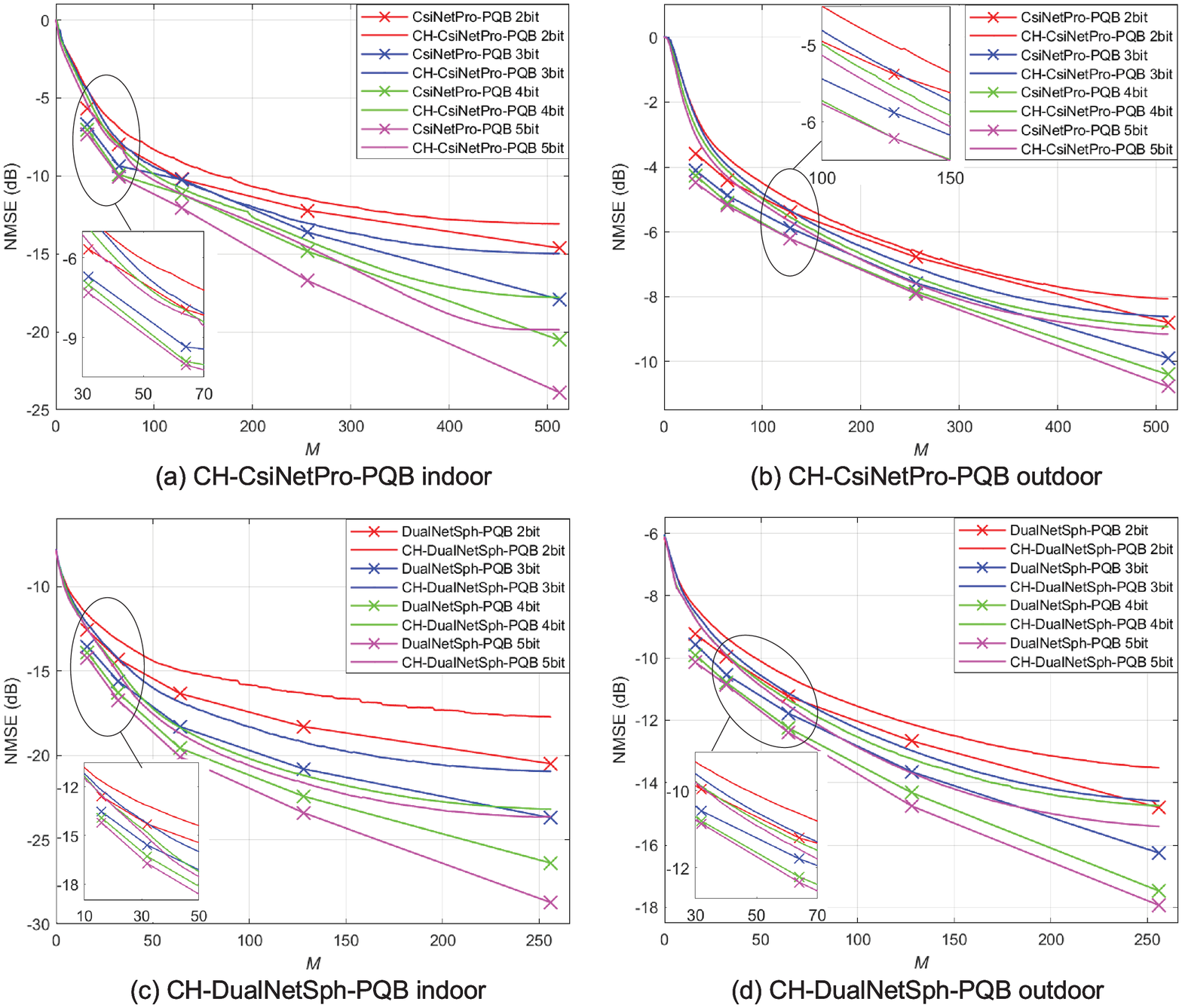}}
  \caption{The NMSE performance of neural networks with the combination of FOCU and PQB with different quantization bits. Total feedback overhead is $M\times b$ bits.}
  \label{fig11} 
\end{figure}

This work further analyzes the above results via the information entropy. The NMSE performance can represent the ability of neural networks to extract codewords information. Therefore, if the entropy of the value of codeword is low, the NMSE performance of neural network will be poor. But if the entropy of the value of codeword is relatively high, the NMSE performance of neural network is depended on the ability of neural network to decode from the codeword. The empirical entropy (bits/element) of codewords under different quantization methods is listed in Table \ref{table5}. For the end-to-end quantization approaches, the codewords of the passing gradient and soft-to-hard quantization have smaller entropy values than that of PQB, which may, to some extent, explain the reason why they have worse performance than PQB. For the $\mu$-law quantization involving retraining, the entropy value of the codewords is greater than that of all of the end-to-end quantization methods. However, the performance of PQB is still better than $\mu$Q. Perhaps this is because the neural network with PQB is optimized in an end-to-end way, and such a global optimization process makes it possible for the decoder to extract much more useful information from the quantized codewords. Although the codewords of $\mu$Q turn out to be more uniformly distributed, the decoder cannot completely exploits the information due to the fact that the neural network is deployed with a sub-optimal retraining process.

\subsection{The Combination of FOCU and PQB}
The proposed changeable-rate and quantization modules are compatible with other DL-based CSI feedback neural networks. Therefore, the neural network can achieve changeable-rate CSI feedback and end-to-end quantization using FOCU and PQB at the same time to jointly improve the efficiency of CSI feedback system.

The NMSE performance of CsiNetPro and DualNetSph with FOCU and PQB, called CH-CsiNetPro-PQB and CH-DualNetSph-PQB are evaluated at the same time. The numbers of the quantization bits $b$ are set from $2$ to $5$. The maximum feedback codeword length is $512$ for CH-CsiNetPro and $256$ for CH-DualNetSph. The lengths $M$ of feedback codewords for CsiNetPro-PQB are $32$, $64$, $128$, $256$, $512$ and for DualNetSph-PQB are $16$, $32$, $64$, $128$, $256$. The total number of feedback bits is $M\times b$.

As shown in Fig. \ref{fig11} (a) to (d), the FOCU can cooperate well with the quantization networks using PQB. The performance of changeable-rate quantization networks is slightly inferior to that of length-fixed of feedback codewords quantization networks with average degradation of about $1.60$ dB, $0.90$ dB and $1.84$ dB and $1.12$ dB, respectively, for CsiNetPro-PQB indoor, CsiNetPro-PQB outdoor, DualNetSph-PQB indoor and DualNetSph-PQB outdoor. It is obviously that when the length of feedback codewords increases, the performance gap between changeable-rate quantization networks and fixed feedback overhead quantization networks increases as well. That is because when the length of feedback codewords increases, CsiNetPro-PQB and DualNetSph-PQB can describe the features more finely to provide the performance gain. However, rather than expressing the specific features for each feedback overhead, changeable-rate networks focus on the common features which exist in all the feedback codewords lengths. In addition, when the number of quantization bits decreases, the performance gap between changeable-rate quantization networks and length-fixed feedback overhead quantization networks becomes smaller.

In summary, this section evaluates and discusses the efficiency and performance improvement of DL-based CSI feedback systems with FOCU and PQB. FOCU can extract the common features of different feedback overhead, thereby it improves the storage efficiency of both the BS and the UE. PQB improves the encoding efficiency of codewords to greatly save the feedback bandwidth. It takes the behavior of the quantization operation into consideration during both forward and back-propagation processes and optimizes the quantization neural networks in an end-to-end way to achieve a better performance with a less codewords entropy compared with other traditional quantization methods.
\section{Conclusion}
DL-based methods are demonstrated to be a promising way in FDD wireless system CSI feedback. This motivates a further research for the improvement of system efficiency. We propose a general DL-based changeable-rate CSI feedback framework with novel quantization operation to improve the efficiency of massive MIMO CSI feedback systems. The proposed CH-CsiNetPro and CH-DualNetSph reutilize all the network layers to achieve overhead-changeable CSI feedback. They save storage space by about 50\% at the UE and the BS sides and do not increase the computation overhead compared with the conventional DL-based length-fixed CSI feedback approaches. The proposed PQB not only improves the encoding efficiency but also has better CSI reconstruction accuracy compared with existing quantization methods.

For future works, it is interesting to exploit the correlation of codewords generated by FOCU to improve the reconstruction accuracy of CSI and utilize the entropy encoding \cite{ref32} to further improve the encoding efficiency of PQB.

\section*{Acknowledgment}
The authors would like to thank the editor and three anonymous reviewers for their valuable suggestions and comments.
The authors would like to thank the helpful discussion with Prof. Zhuqing Jia at Beijing University of Posts and Telecommunications.

\bibliographystyle{IEEEtran}
\bibliography{ref}

\begin{thebibliography}{10}
\providecommand{\url}[1]{#1}
\csname url@samestyle\endcsname
\providecommand{\newblock}{\relax}
\providecommand{\bibinfo}[2]{#2}
\providecommand{\BIBentrySTDinterwordspacing}{\spaceskip=0pt\relax}
\providecommand{\BIBentryALTinterwordstretchfactor}{4}
\providecommand{\BIBentryALTinterwordspacing}{\spaceskip=\fontdimen2\font plus
\BIBentryALTinterwordstretchfactor\fontdimen3\font minus
  \fontdimen4\font\relax}
\providecommand{\BIBforeignlanguage}[2]{{%
\expandafter\ifx\csname l@#1\endcsname\relax
\typeout{** WARNING: IEEEtran.bst: No hyphenation pattern has been}%
\typeout{** loaded for the language `#1'. Using the pattern for}%
\typeout{** the default language instead.}%
\else
\language=\csname l@#1\endcsname
\fi
#2}}
\providecommand{\BIBdecl}{\relax}
\BIBdecl

\bibitem{ref1}
Y.~Liu, X.~Wang, G.~Boudreau, A.~B. Sediq, and H.~Abou-Zeid, ``A
  multi-dimensional intelligent multiple access technique for {5G} beyond and
  {6G} wireless networks,'' \emph{IEEE Transactions on Wireless
  Communications}, vol.~20, no.~2, pp. 1308--1320, 2021.

\bibitem{ref2}
C.~{Oestges}, ``Overview of radio propagation models for {5G} and beyond,'' in
  \emph{2019 URSI Asia-Pacific Radio Science Conference (AP-RASC)}, 2019, pp.
  1--1.

\bibitem{ref3}
B.~{Zong}, X.~{Duan}, C.~{Fan}, and K.~{Guan}, ``{6G} technologies -
  opportunities and challenges,'' in \emph{2020 IEEE International Conference
  on Integrated Circuits, Technologies and Applications (ICTA)}, 2020, pp.
  171--173.

\bibitem{ref4}
I.~F. {Akyildiz}, A.~{Kak}, and S.~{Nie}, ``{6G} and beyond: The future of
  wireless communications systems,'' \emph{IEEE Access}, vol.~8, pp.
  133\,995--134\,030, 2020.

\bibitem{ref5}
W.~{Hong}, Z.~H. {Jiang}, C.~{Yu}, J.~{Zhou}, P.~{Chen}, Z.~{Yu}, H.~{Zhang},
  B.~{Yang}, X.~{Pang}, M.~{Jiang}, Y.~{Cheng}, M.~K.~T. {Al-Nuaimi},
  Y.~{Zhang}, J.~{Chen}, and S.~{He}, ``Multibeam antenna technologies for {5G}
  wireless communications,'' \emph{IEEE Transactions on Antennas and
  Propagation}, vol.~65, no.~12, pp. 6231--6249, 2017.

\bibitem{ref6}
J.~Yuan, H.~Q. Ngo, and M.~Matthaiou, ``Machine learning-based channel
  prediction in massive {MIMO} with channel aging,'' \emph{IEEE Transactions on
  Wireless Communications}, vol.~19, no.~5, pp. 2960--2973, 2020.

\bibitem{ref7}
M.~{Barzegar Khalilsarai}, S.~{Haghighatshoar}, and G.~{Caire}, ``How to
  achieve massive {MIMO} gains in {FDD} systems?'' in \emph{2018 IEEE 19th
  International Workshop on Signal Processing Advances in Wireless
  Communications (SPAWC)}, 2018, pp. 1--5.

\bibitem{ref8}
3GPP, ``{NR; Physical layer procedures for data},'' {3rd Generation Partnership
  Project (3GPP)}, Technical Specification (TS) 38.214, Mar 2020, version
  16.1.0.

\bibitem{ref9}
P.~{Kuo}, H.~T. {Kung}, and P.~{Ting}, ``Compressive sensing based channel
  feedback protocols for spatially-correlated massive antenna arrays,'' in
  \emph{2012 IEEE Wireless Communications and Networking Conference (WCNC)},
  2012, pp. 492--497.

\bibitem{ref10}
X.~{Rao} and V.~K.~N. {Lau}, ``Distributed compressive {CSIT} estimation and
  feedback for {FDD} multi-user massive {MIMO} systems,'' \emph{IEEE
  Transactions on Signal Processing}, vol.~62, no.~12, pp. 3261--3271, 2014.

\bibitem{ref11}
D.~{Li} and Y.~{Dong}, ``Deep learning: methods and applications,''
  \emph{Foundations and trends in signal processing}, vol.~7, no. 3--4, pp.
  197--387, 2014.

\bibitem{ref12}
J.~{Zhang}, M.~O. {Akinsolu}, B.~{Liu}, and G.~A.~E. {Vandenbosch}, ``Automatic
  {AI}-driven design of mutual coupling reducing topologies for frequency
  reconfigurable antenna arrays,'' \emph{IEEE Transactions on Antennas and
  Propagation}, vol.~69, no.~3, pp. 1831--1836, 2021.

\bibitem{ref13}
M.~{Wu} and L.~{Chen}, ``Image recognition based on deep learning,'' in
  \emph{2015 Chinese Automation Congress (CAC)}, 2015, pp. 542--546.

\bibitem{ref14}
W.~{Wang}, Y.~{Huang}, Y.~{Wang}, and L.~{Wang}, ``Generalized autoencoder: A
  neural network framework for dimensionality reduction,'' in \emph{2014 IEEE
  Conference on Computer Vision and Pattern Recognition Workshops}, 2014, pp.
  496--503.

\bibitem{ref15}
D.~{Wang}, J.~{Su}, and H.~{Yu}, ``Feature extraction and analysis of natural
  language processing for deep learning {English} language,'' \emph{IEEE
  Access}, vol.~8, pp. 46\,335--46\,345, 2020.

\bibitem{ref16}
S.~{Ma}, Y.~{Zhu}, G.~{Wang}, and R.~{He}, ``Machine learning aided channel
  estimation for ambient backscatter communication systems,'' in \emph{2018
  IEEE International Conference on Communication Systems (ICCS)}, 2018, pp.
  67--71.

\bibitem{ref17}
Z.~{Mao} and S.~{Yan}, ``Deep learning based channel estimation in fog radio
  access networks,'' \emph{China Communications}, vol.~16, no.~11, pp. 16--28,
  2019.

\bibitem{ref18}
C.~{Wen}, W.~{Shih}, and S.~{Jin}, ``Deep learning for massive {MIMO} {CSI}
  feedback,'' \emph{IEEE Wireless Communications Letters}, vol.~7, no.~5, pp.
  748--751, 2018.

\bibitem{ref19}
Z.~{Liu}, L.~{Zhang}, and Z.~{Ding}, ``Exploiting bi-directional channel
  reciprocity in deep learning for low rate massive {MIMO} {CSI} feedback,''
  \emph{IEEE Wireless Communications Letters}, vol.~8, no.~3, pp. 889--892,
  2019.

\bibitem{ref20}
T.~{Wang}, C.~{Wen}, S.~{Jin}, and G.~Y. {Li}, ``Deep learning-based {CSI}
  feedback approach for time-varying massive {MIMO} channels,'' \emph{IEEE
  Wireless Communications Letters}, vol.~8, no.~2, pp. 416--419, 2019.

\bibitem{ref21}
\BIBentryALTinterwordspacing
J.~{Guo}, X.~{Yang}, C.~{Wen}, S.~{Jin}, and G.~Y. {Li}, ``{DL}-based {CSI}
  feedback and cooperative recovery in massive {MIMO},'' \emph{arXiv preprint
  arXiv:2003.03303}, 2020. [Online]. Available: \url{[online]
  https://arxiv.org/abs/2003.03303}
\BIBentrySTDinterwordspacing

\bibitem{ref22}
J.~{Guo}, C.~{Wen}, S.~{Jin}, and G.~Y. {Li}, ``Convolutional neural
  network-based multiple-rate compressive sensing for massive {MIMO} {CSI}
  feedback: Design, simulation, and analysis,'' \emph{IEEE Transactions on
  Wireless Communications}, vol.~19, no.~4, pp. 2827--2840, 2020.

\bibitem{ref23}
Z.~{Liu}, M.~{del Rosario}, X.~{Liang}, L.~{Zhang}, and Z.~{Ding}, ``Spherical
  normalization for learned compressive feedback in massive {MIMO} {CSI}
  acquisition,'' in \emph{2020 IEEE International Conference on Communications
  Workshops (ICC Workshops)}.\hskip 1em plus 0.5em minus 0.4em\relax IEEE,
  2020, pp. 1--6.

\bibitem{ref24}
H.~{Ye}, G.~Y. {Li}, and B.~{Juang}, ``Power of deep learning for channel
  estimation and signal detection in {OFDM} systems,'' \emph{IEEE Wireless
  Communications Letters}, vol.~7, no.~1, pp. 114--117, 2018.

\bibitem{ref25}
H.~{Ye}, L.~{Liang}, G.~Y. {Li}, and B.~{Juang}, ``Deep learning-based
  end-to-end wireless communication systems with conditional {GANs} as unknown
  channels,'' \emph{IEEE Transactions on Wireless Communications}, vol.~19,
  no.~5, pp. 3133--3143, 2020.

\bibitem{ref26}
C.~{Huang}, A.~F. {Molisch}, R.~{He}, R.~{Wang}, P.~{Tang}, B.~{Ai}, and
  Z.~{Zhong}, ``Machine learning-enabled {LOS/NLOS} identification for {MIMO}
  systems in dynamic environments,'' \emph{IEEE Transactions on Wireless
  Communications}, vol.~19, no.~6, pp. 3643--3657, 2020.

\bibitem{ref27}
J.~{Guo}, X.~{Li}, M.~{Chen}, P.~{Jiang}, T.~{Yang}, W.~{Duan}, H.~{Wang},
  S.~{Jin}, and Q.~{Yu}, ``{AI} enabled wireless communications with real
  channel measurements: Channel feedback,'' \emph{Journal of Communications and
  Information Networks}, vol.~5, no.~3, pp. 310--317, 2020.

\bibitem{ref28}
L.~{Liu}, C.~{Oestges}, J.~{Poutanen}, K.~{Haneda}, P.~{Vainikainen},
  F.~{Quitin}, F.~{Tufvesson}, and P.~D. {Doncker}, ``The {COST} 2100 {MIMO}
  channel model,'' \emph{IEEE Wireless Communications}, vol.~19, no.~6, pp.
  92--99, 2012.

\bibitem{ref29}
Y.~{Jang}, G.~{Kong}, M.~{Jung}, S.~{Choi}, and I.~{Kim}, ``Deep autoencoder
  based {CSI} feedback with feedback errors and feedback delay in {FDD} massive
  {MIMO} systems,'' \emph{IEEE Wireless Communications Letters}, vol.~8, no.~3,
  pp. 833--836, 2019.

\bibitem{ref30}
T.~{Chen}, J.~{Guo}, S.~{Jin}, C.~{Wen}, and G.~Y. {Li}, ``A novel quantization
  method for deep learning-based massive {MIMO} {CSI} feedback,'' in \emph{2019
  IEEE Global Conference on Signal and Information Processing (GlobalSIP)},
  2019, pp. 1--5.

\bibitem{ref31}
D.~Kingma and J.~Ba, ``Adam: A method for stochastic optimization,''
  \emph{Computer Science}, 2014.

\bibitem{ref32}
M.~B. Mashhadi, Q.~Yang, and D.~Gündüz, ``Distributed deep convolutional
  compression for massive {MIMO} {CSI} feedback,'' \emph{IEEE Transactions on
  Wireless Communications}, vol.~20, no.~4, pp. 2621--2633, 2021.

\bibitem{deep_tasks}
\BIBentryALTinterwordspacing
N.~Shlezinger and Y.~C. Eldar, ``Deep task-based quantization,''
  \emph{Entropy}, vol.~23, no.~1, 2021. [Online]. Available:
  \url{https://www.mdpi.com/1099-4300/23/1/104}
\BIBentrySTDinterwordspacing

\bibitem{WalterRudin0Functional}
W.~Rudin, \emph{Functional analysis / 2nd ed}.\hskip 1em plus 0.5em minus
  0.4em\relax Singapore: McGraw-Hill Book Company, 1991.

\bibitem{SALDR}
X.~Song, J.~Wang, J.~Wang, G.~Gui, T.~Ohtsuki, H.~Gacanin, and H.~Sari,
  ``{SALDR}: Joint self-attention learning and dense refine for massive {MIMO}
  {CSI} feedback with multiple compression ratio,'' \emph{IEEE Wireless
  Communications Letters}, vol.~10, no.~9, pp. 1899--1903, 2021.

\end{thebibliography}
\ifCLASSOPTIONcaptionsoff
  \newpage
\fi

\end{document}